\documentclass[oneside, reqno, 10pt, a4paper]{amsart}

\title[Opal.jl: a comprehensive, composable framework for data assimilation in Julia]{Opal.jl: a comprehensive, composable framework for data assimilation in Julia}
\date{\today}
\address{$^\dagger$Faculty Of Civil Engineering and Geosciences\\TU Delft\\Delft\\The Netherlands}
\author[N. Mueller]{Nicholas Mueller$^\dagger$}
\email{nmueller@tudelft.nl}

\usepackage{microtype,times}
\AtBeginDocument{
  \DeclareSymbolFont{AMSb}{U}{msb}{m}{n}
  \DeclareSymbolFontAlphabet{\mathbb}{AMSb}
}
\DeclareFontFamily{U}{mathx}{\hyphenchar\font45}
\DeclareFontShape{U}{mathx}{m}{n}{<-> mathx10}{}
\DeclareSymbolFont{mathx}{U}{mathx}{m}{n}
\DeclareMathAccent{\widebar}{0}{mathx}{"73}
\DeclareMathAccent{\widecheck}{0}{mathx}{"71}

\usepackage[dvipsnames]{xcolor}
\usepackage{graphicx}
\usepackage{subfig}
\usepackage{svg}
\usepackage{tikz}
\tikzstyle{arrow} = [thick,->,>=stealth]
\usetikzlibrary{shapes.misc,matrix,fit,positioning,arrows.meta,decorations.pathreplacing,calc,shapes.geometric,arrows,trees}
\usepackage{forest}
\forestset{
  default preamble={
    for tree={
      draw, thick, rounded corners=3pt, align=center,
      font=\scriptsize\fontfamily{qcr}\selectfont, l sep=10mm, s sep=4mm,
      edge={draw=gray,thick}, parent anchor=south, child anchor=north,
      inner sep=3pt,
    }
  },
  abs/.style={draw=gray,fill=myred!25},
  conc/.style={draw=gray,fill=myblue!25},
  wrap/.style={draw=gray,fill=mydarkblue!25},
  alias/.style={draw=gray,fill=myorange!25},
}
\tikzset{Matrix/.style={matrix of nodes, font=\footnotesize,text height=1pt, text depth=0.5pt, text width=8.5pt, align=center, column sep=0pt, row sep=0pt, nodes in empty cells}}
\tikzset{cross/.style={cross out, draw=black, minimum size=2*(#1-\pgflinewidth), inner sep=0pt, outer sep=0pt},cross/.default={3pt}}
\usepackage[a4paper, pdftex, left=2cm, top=2cm, right=2cm, bottom=2cm]{geometry}
\usepackage[final]{pdfpages}
\usepackage{changepage}

\usepackage[foot]{amsaddr}

\usepackage{url}
\usepackage{hyperref}
\hypersetup{
    breaklinks=true,
    bookmarksopen=true,
    pdftitle={Opal.jl: a composable, model-error-aware framework for sequential and variational data assimilation in Julia},
    pdfauthor={Nicholas Mueller},
    pdfsubject={Data assimilation, uncertainty quantification, Kalman filtering, reduced order modelling, Julia},
    pdfkeywords={data assimilation, ensemble Kalman filter, reduced order modelling, bias-aware filtering, Julia},
    colorlinks=true,
    linkcolor=black,
    citecolor=blue,
    filecolor=black,
    urlcolor=blue
}

\usepackage[numbers,sort&compress]{natbib}

\PassOptionsToPackage{normalem}{ulem}
\usepackage{ulem}
\providecolor{added}{rgb}{0,0,1}
\providecolor{deleted}{rgb}{1,0,0}

\usepackage{float}
\usepackage{framed}
\usepackage{verbatim}
\usepackage{fancyvrb}
\usepackage{booktabs}
\usepackage{colortbl}
\usepackage{multirow}
\usepackage{algorithm}
\usepackage{algpseudocode}
\makeatletter
\algdef{S}[IF]{IfNoThen}[1]{\algorithmicif\ #1}
\makeatother
\usepackage[inline]{enumitem}
\usepackage{siunitx}
\usepackage{mathtools}
\mathtoolsset{showonlyrefs}

\newtheorem{remark}{Remark}

\usepackage[breakable]{tcolorbox}
\tcbuselibrary{theorems}
\tcbuselibrary{skins}
\tcbset{
	commonstyle/.style={
		theorem style=plain,
		enhanced jigsaw,
		fonttitle=\bfseries,
		fontupper=\itshape,
		halign=justify,
		separator sign=:,
		description delimiters none,
		description font=\bfseries, 
		terminator sign={.\hspace{0.25em}},
		arc=0mm,outer arc=0mm,
		boxrule=0pt,toprule=0pt,bottomrule=0pt,leftrule=0pt,rightrule=0pt,
		titlerule=0pt,toptitle=0pt,bottomtitle=0pt,top=0pt,
		colback=white,coltitle=black,
		boxsep=0pt, bottom=0pt, left=0pt, 
	}
}
\newtcbtheorem[]{myproblem}{Problem}%
{center, commonstyle, fonttitle=\bfseries}{pb}
\newtcbtheorem[]{mydefinition}{Definition}
{center, commonstyle, fonttitle=\bfseries}{pb}
\newtcbtheorem[]{myassumption}{Assumption}
{center, commonstyle, fonttitle=\bfseries}{pb}

\definecolor{myred}{RGB}{235,87,87}         
\definecolor{myblue}{RGB}{52,152,219}       
\definecolor{mygreen}{RGB}{46,204,113}      
\definecolor{myorange}{RGB}{243,156,18}     
\definecolor{mydarkblue}{RGB}{155,89,182}   
\definecolor{plotblue}{HTML}{1f77b4}
\definecolor{plotred}{HTML}{d62728}
\definecolor{plotgreen}{HTML}{2ca02c}
\definecolor{plotpurple}{HTML}{9467bd}
\newcommand{\legenddot}[1]{\tikz[baseline=-0.5ex]{\fill[#1] (0,0) circle (0.6ex);}}

\usepackage{amsmath, amsfonts, amssymb, amscd, bm, bbm, mathtools, stmaryrd}

\newcommand{\R}{\mathbb{R}}

\usepackage{acronym}
\acrodef{ode}[ODE]{ordinary differential equation}
\acrodef{pde}[PDE]{partial differential equation}
\acrodef{fe}[FE]{finite element}
\acrodef{dg}[DG]{discontinuous Galerkin}
\acrodef{cdg}[C/DG]{continuous/discontinuous Galerkin}
\acrodef{fem}[FEM]{finite element method}
\acrodef{bem}[BEM]{boundary element method}
\acrodef{dof}[DOF]{degree of freedom}
\acrodefplural{dof}[DOFs]{degrees of freedom}
\acrodef{be}[BE]{Backward Euler}
\acrodef{hf}[HF]{high fidelity}
\acrodef{fom}[FOM]{full order model}
\acrodef{lhs}[LHS]{left hand side}
\acrodef{rhs}[RHS]{right hand side}
\acrodef{rom}[ROM]{reduced order model}
\acrodef{rb}[RB]{reduced basis}
\acrodefplural{rb}[RBs]{reduced bases}
\acrodef{svd}[SVD]{singular value decomposition}
\acrodef{sthosvd}[ST-HOSVD]{sequentially truncated high order singular value decomposition}
\acrodef{pod}[POD]{proper orthogonal decomposition}
\acrodef{tpod}[TPOD]{truncated proper orthogonal decomposition}
\acrodef{strb}[ST-RB]{space-time reduced basis}
\acrodef{eim}[EIM]{empirical interpolation method}
\acrodef{deim}[DEIM]{discrete empirical interpolation method}
\acrodef{mdeim}[MDEIM]{its matrix counterpart}
\acrodef{stmdeimrb}[ST-MDEIM-RB]{space-time MDEIM-RB}
\acrodef{dl}[DL]{deep learning}
\acrodef{nn}[NN]{neural network}
\acrodef{tt}[TT]{tensor-train}
\acrodef{ttrb}[TT-RB]{tensor-train reduced basis}
\acrodef{podrb}[POD-RB]{proper orthogonal decomposition reduced basis}
\acrodef{ttsvd}[TT-SVD]{tensor-train SVD}
\acrodef{ttcross}[TT-CROSS]{tensor-train cross}
\acrodef{ttmdeim}[TT-MDEIM]{tensor-train MDEIM}
\acrodef{jit}[JIT]{just-in-time}
\acrodef{rbf}[RBF]{radial basis function}
\acrodefplural{rbf}[RBFs]{radial bases functions}
\acrodef{ale}[ALE]{arbitrary Eulerian-Lagriangian}
\acrodef{fsi}[FSI]{fluid-structure interaction}
\acrodef{vlfs}[VLFS]{very large floating structure}
\acrodef{uq}[UQ]{uncertainty quantification}
\acrodef{da}[DA]{Data assimilation}
\acrodef{kf}[KF]{Kalman filter}
\acrodef{ekf}[EKF]{extended Kalman filter}
\acrodef{enkf}[EnKF]{ensemble Kalman filter}
\acrodef{denkf}[DEnKF]{deterministic ensemble Kalman filter}
\acrodef{ukf}[UKF]{unscented Kalman filter}
\acrodef{nll}[NLL]{negative log-likelihood}
\acrodef{mle}[MLE]{maximum likelihood estimate}
\acrodef{rc}[RC]{reservoir computing}
\acrodef{esn}[ESN]{echo state network}
\acrodef{rem}[REM]{reduced error model}
\acrodef{blup}[BLUP]{best linear unbiased prediction}
\acrodef{mc}[MC]{Monte Carlo}
\acrodef{mcmc}[MCMC]{Markov chain Monte Carlo}
\acrodef{pdf}[PDF]{probability density function}
\acrodef{sir}[SIR]{sequential importance resampling}

\graphicspath{{./}}

\newcommand{\fig}[1]{Figure~\ref{#1}}
\newcommand{\tab}[1]{Table~\ref{#1}}

\newcommand{\lst}[1]{Listing~\ref{#1}}

\newcommand{\myfont}[1]{{\fontfamily{qcr}\selectfont #1}}

\begin{document}

\maketitle

\begin{abstract}
Data assimilation (DA) and inverse modelling are indispensable tools for combining physics-based models with observations, yet the software that implements them is often fragmented: sequential (Kalman and particle) filters and variational (3D/4D-Var) estimators are typically developed as separate codebases, each tied to a specific class of forward model -- for example, simple time-marching schemes for ordinary differential equations (ODEs), or full finite element (FE) discretisations of partial differential equations (PDEs). In this work, we present \textsc{Opal.jl}, a Julia package that overcomes this fragmentation by providing a unified environment for a wide range of DA methods and forward model backends, all accessible behind a single high-level interface. Starting from a handful of basic methods, complex inference tools can be built by composing them with advanced capabilities, such as covariance localisation/inflation, online noise covariance estimation, bias-aware correction, and kriging-based calibration of reduced-order surrogates. The package integrates natively with the SciML ecosystem for ODE-governed systems, and with Gridap/GridapROMs for both full-order and reduced-order discretisations of PDEs. Notably, the API is designed to allow a reduced-order surrogate to replace a full-order solver with no change to the code, even at the driver level. We demonstrate the library on a Lorenz-63 benchmark, on a Van der Pol oscillator problem with biased observations, on a turbulent Navier-Stokes flow past a square cavity, where the unknown Reynolds number is inferred from sparse velocity and pressure measurements, and finally on a heat equation using a reduced-order surrogate instead of a more standard full-order model.
\end{abstract}
\noindent
{\bf Program summary} \\
{\em Program Title:} Opal.jl: Open source Probabilistic \& Assimilation Library in Julia \\
{\em Program Version:} 0.1.0 \\
{\em Developer's repository link:} https://github.com/nichomueller/Opal.jl \\
{\em Licensing provisions:} MIT license \\
{\em Programming language:} Julia \\
{\em Nature of problem:} Estimating the hidden state and, where relevant, the uncertain parameters of any dynamical system from observational data that are partial, corrupted by noise, and potentially biased. Such problems are generally ill-posed and computationally demanding, since they usually rely on a forecast-and-analyse paradigm, with the forecasting step being potentially expensive. Practitioners typically lack a single and mature software environment across heterogeneous forward model backends -- lightweight ODEs and full-order PDE discretisations alike -- without extensive re-implementation. \\
{\em Solution method:} Opal.jl exposes a single, type-driven interface to both Bayesian sequential filters (Kalman, extended Kalman, unscented, ensemble Kalman variants, and particle filters) and variational methods (3D-Var, 4D-Var), all sharing the same execution, composition, and diagnostic pipeline. Advanced capabilities -- covariance localisation and inflation, surrogate model calibration, and bias-aware correction -- are added incrementally as composable wrappers around a common filter interface. The library integrates natively with the SciML ecosystem for ODE-governed models and with Gridap/GridapROMs for both full-order and reduced-order PDE models, thus enabling a wide range of applications to be tackled with multiple methodologies all sharing the same pipeline. \\

\section{Introduction}
\label{sec:introduction}

The solution of inverse problems is a fundamental task across the sciences and engineering, including meteorology \cite{Daley1991,Bennett2002}, oceanography \cite{Wunsch1996,Rodgers2000}, hydrology \cite{Anderson2015}, and geophysics \cite{Menke2018}. Inverse problems seek to estimate unknown parameters and/or unobserved states of a system from data measurements that are partial and corrupted by noise. Because such problems are typically ill-posed \cite{manzoni2016accurate}, their robust and efficient solution is essential for reliable forecasting.

\ac{da} is a particularly well-developed instance of this general problem, in which a forward model (often high-dimensional and nonlinear), describing some physical process, is combined with observations to estimate an evolving hidden state and, in many cases, uncertain model parameters. Two methodological families have matured largely in parallel: sequential Bayesian filters -- among which \acp{kf} \cite{Kalman1960} and particle filters \cite{doucet2001sequential} are especially popular -- and variational methods, such as the 3D- and 4D-Var \cite{lorenc1986analysis,talagrand1987variational}. Despite the two families being conceptually overlapping, the corresponding software has remained comparatively fragmented: the two approaches are rarely available together in open-source packages and are typically decoupled from the forward model backend, so that switching from a hand-written prototype to a dynamical system based on the \ac{fe} discretisation of a \ac{pde} requires substantial re-engineering. In practice, this forces users to reimplement even straightforward routines such as observation handling, ensemble bookkeeping, and diagnostics for every new combination of method and model, slowing methodological comparison and greatly reducing productivity.

The Julia ecosystem already provides mature, well-established packages that individually address parts of this landscape, though none spans it entirely. EnsembleKalmanProcesses.jl~\cite{EnsembleKalmanProcesses} implements \ac{enkf}- and \ac{ukf}-based inversion for calibrating black-box models, but it lacks several hallmarks of sequential state-estimation theory and features no backends for \ac{ode}/\ac{pde}-based modelling. LowLevelParticleFilters.jl~\cite{LowLevelParticleFilters} offers an established implementation of the \ac{ekf}, \ac{ukf}, and particle filters for state estimation, but does not extend to variational estimation and suffers from a similarly restricted scope of applications. General-purpose probabilistic-programming packages such as Turing.jl~\cite{Turing} and Gen.jl~\cite{Gen} support Bayesian inference over essentially arbitrary generative models, but are more focused on regression and \ac{mcmc} sampling strategies than on \ac{da}-specific filtering, and have not been applied to large-scale \ac{pde}-constrained state estimation. Within the SciML ecosystem itself, DiffEqParamEstim.jl and SciMLSensitivity.jl~\cite{SciMLSensitivity} provide optimisation- and adjoint-based parameter fitting for \acp{ode}, addressing deterministic parameter estimation but not Bayesian inference, advanced bias-correction capabilities, or \ac{pde} modelling. On the other hand, \ac{uq}-oriented packages such as UncertaintyQuantification.jl~\cite{UncertaintyQuantification} and GlobalSensitivity.jl~\cite{GlobalSensitivity} provide \ac{mc}, polynomial-chaos, and sensitivity-analysis methods for propagating uncertainty through a model, but are not organised around a \ac{da} loop. Additionally, none of the above address the need for combining \ac{da} with \acp{rom}, whose use is paramount for large-scale \ac{pde}-based applications.

Against this background, Opal addresses the lack of a unified and extensible framework for \ac{da} and inverse modelling by bringing together modern algorithms for sequential filtering and variational estimation across a broad range of forward models. Several of these capabilities are currently absent from existing Julia frameworks and are rarely available together within a single computational ecosystem. Leveraging Julia's multiple dispatch and \ac{jit} compilation, Opal combines efficiency and flexibility while exposing these methods through a common interface applicable to forward models ranging from arbitrary \ac{ode} systems to \ac{fe}-discretised \acp{pde} and their \ac{rom} counterparts.

A second contribution is a compositional design that avoids embedding advanced capabilities into monolithic algorithm implementations. Instead, Opal expresses extensions as wrappers around a shared filter interface: covariance localisation~\cite{hamill2001distance}, multiplicative and innovation-adaptive covariance inflation \cite{sun2024high}, adaptive noise covariance estimation \cite{berry2013adaptive}, kriging-based calibration of reduced-order surrogates \cite{pagani2017efficient}, and bias correction through \ac{rc}-based architectures~\cite{Novoa2023} can each be incorporated incrementally around an existing filter. New filters can be added either by combining existing wrappers or by implementing the interface directly, without changes to the existing codebase.

A third contribution is interoperability with established scientific-computing ecosystems. For \ac{ode}-based dynamics, a thin layer allows us to use arbitrary SciML integrators directly as forward models. For \ac{pde}-based dynamics, Opal interfaces with Gridap~\cite{Badia2020} and GridapROMs~\cite{muellerbadiagridaproms}, enabling the solutions -- either full-order or reduced-order -- to be assimilated by any filtering method in the library without modification of the assimilation workflow beyond replacing the transition model. Consequently, a cheap surrogate can replace the corresponding full-order model purely as a computational acceleration. This integration connects \ac{da} methodology with modern discretised physics models, overcoming software boundaries that traditionally limit the transfer of methods between algorithm development and simulation frameworks operating at different fidelities.

The paper is organised as follows. Section~\ref{sec:description} formalises the estimation problem and introduces the methodologies implemented in Opal. Section~\ref{sec:implementation} describes the package architecture, composable wrapper mechanism, SciML and Gridap/GridapROMs interoperability layers, and a fully commented usage example. Section~\ref{sec:examples} demonstrates the capabilities of the framework on four challenging benchmarks spanning chaotic \ac{ode} and \ac{pde} dynamics. First, a Lorenz-63 system is addressed with multiple different Bayesian and variational methods. Second, a Van der Pol oscillator problem with biased observations is solved with a \ac{rc}-corrected, bias-aware ensemble filter. Third, a turbulent two-dimensional Navier-Stokes flow around a square cavity showcases joint state-parameter estimation through the identification of the unknown Reynolds number via \ac{da}. Lastly, a parametrised heat equation demonstrates the replacement of a full-order \ac{pde} solver by a reduced-order surrogate, with kriging-based calibration compensating for the associated loss of accuracy. Code snippets for the usage example in Section~\ref{sec:implementation} and the numerical experiments are reported throughout the body of the paper. Finally, Section~\ref{sec:conclusions} concludes the paper with a summary of the content and future directions of research.
\section{Problem description and methods}
\label{sec:description}

This section lays the mathematical foundations of the problems addressed by Opal. Subsection~\ref{subs:model} formalises the class of joint state-parameter inverse models targeted by the package. Subsection~\ref{subs:kf} and~\ref{subs:enkf} derive the classical \ac{kf} and its ensemble variants as the two basic instances of this framework, while Subsection~\ref{subs:infl_loc_adapt} discusses the auxiliary inflation, localisation, and adaptivity techniques that increase the stability and accuracy of the \acp{enkf} across different applications. Subsections~\ref{subs:bias-aware} and~\ref{subs:rem} extend the filter to account for two distinct sources of structural model error -- systematic observation bias and the error introduced by a reduced-order surrogate -- through a \ac{rc}-based and a kriging-calibrated bias model, respectively. Subsection~\ref{subs:misc} provides a brief overview of the remaining Bayesian methodologies implemented in the package (the \ac{ukf} and particle filters), while Subsection~\ref{subs:variational} closes the section by discussing variational (3D-/4D-Var) methods.

\subsection{The model}
\label{subs:model}

By and large, we are interested in the following class of inverse problems: given a set of observations of a system governed by a parameterised system of equations, we aim to jointly recover an unknown state variable and a parameter vector that best explain the observations. This problem is described by a constrained optimisation problem
\begin{equation}
  \begin{aligned}
    \label{eq: opt problem deterministic}
    \arg\min_{(\bm{\mu},\bm{u})} \;\mathcal{L}(\bm{u}(\bm{\mu});\bm{\mu})
    &\doteq \frac12 \left\| \bm{\mathcal{G}}(\bm{u}(\bm{\mu})) - \bm{u}_{\mathrm{obs}} \right\|_{\bm{\Sigma}^{-1}}^2, \\
    \text{subject to} \quad
    \frac{d\bm{u}(\bm{\mu})}{dt} + \bm{\mathcal{A}}(\bm{u}(\bm{\mu});\bm{\mu}) &= \bm{0},
  \end{aligned}
\end{equation}
\noindent
where $\bm{u}(\bm{\mu}) \in \R^{N_u}$ is the state variable, $\bm{\mu} \in \R^{N_{\mu}}$ is the parameter vector, $\bm{\mathcal{G}}: \R^{N_u} \to \R^{N_{\mathrm{obs}}}$ is the \emph{observation} operator, $\bm{u}_{\mathrm{obs}} \in \R^{N_{\mathrm{obs}}}$ are the observations, and $\bm{\Sigma} \in \R^{N_{\mathrm{obs}} \times N_{\mathrm{obs}}}$ is a symmetric, positive-definite weighting matrix specified shortly. The operator $\bm{\mathcal{A}}$ describes the dynamics of the system -- which may be governed, for example, by a set of \acp{ode} or \acp{pde} -- and is commonly referred to as the \emph{forward} or \emph{transition} model. The operator $\bm{\mathcal{G}}$ is usually linear, that is
\begin{equation}
  \label{eq: bias-aware obs}
  \bm{\mathcal{G}}(\bm{u}) = \bm{G}\bm{u} \ \ (+ \bm{b}), \qquad 
  \bm{G} \in \R^{N_{\mathrm{obs}} \times N_u}, \bm{b} \in \R^{N_{\mathrm{obs}}},
\end{equation}
where $\bm{b}$ is an optional bias term, which may be included to account for systematic errors in the observations. There are many examples of bias-aware implementations in the literature, including \textit{in-situ} bias estimation and correction \cite{Rubio2019,daSilvaColonius2020}, and hybrid bias-aware \ac{da} \cite{Chinesta2020,BonavitaLaloyaux2020}. In very simple terms, the former estimates the bias as part of the assimilation framework, whereas the latter decouples these two tasks and is generally credited with greater flexibility and complexity in the bias model \cite{Novoa2023}. The treatment of bias is central to the present work, and we will return to it in Subsections~\ref{subs:bias-aware}-\ref{subs:rem}. \\

\noindent In practice, however, neither $\bm{\mathcal{A}}$ nor $\bm{\mathcal{G}}$ are known exactly. A standard way to represent the uncertainty associated with these operators is by introducing additive Gaussian noise, that is
\begin{equation}
  \begin{aligned}
    \bm{\mathcal{A}}_{\bm{\theta}}(\cdot) \doteq \bm{\mathcal{A}}(\cdot) + \bm{\theta}, 
    \qquad 
    \bm{\mathcal{G}}_{\bm{\sigma}}(\cdot) \doteq \bm{\mathcal{G}}(\cdot) + \bm{\sigma},  
  \end{aligned}
\end{equation}
where $\bm{\theta} \sim \mathcal{N}(\bm{0},\bm{\Theta})$ and $\bm{\sigma} \sim \mathcal{N}(\bm{0},\bm{\Sigma})$ are independent, zero-mean Gaussian random variables. Note that the covariance matrix $\bm{\Sigma}$ for the observation noise corresponds to the weight employed in Eq.~\eqref{eq: opt problem deterministic} for the observation misfit. Following the standard \ac{da} and \ac{uq} literature -- see, e.g., the seminal work of Kalman \cite{Kalman1960} -- the true pair $(\bm{\mu}^{\star},\bm{u}^{\star})$, i.e., the one actually realising the observations through the unknown true system, is assumed to satisfy not Eq.~\eqref{eq: opt problem deterministic} but rather the stochastic problem
\begin{equation}
  \begin{aligned}
    \label{eq: opt problem stochastic}
    \frac{d\bm{u}^{\star}(\bm{\mu}^{\star})}{dt} + \bm{\mathcal{A}}_{\bm{\theta}^{\star}}(\bm{u}^{\star}(\bm{\mu}^{\star});\bm{\mu}^{\star}) &= \bm{0}, \\
    \bm{u}_{\mathrm{obs}} &= \bm{\mathcal{G}}_{\bm{\sigma}^{\star}}(\bm{u}^{\star}(\bm{\mu}^{\star})),
  \end{aligned}
\end{equation}
with $\bm{\theta}^{\star} \sim \mathcal{N}(\bm{0},\bm{\Theta})$ and $\bm{\sigma}^{\star} \sim \mathcal{N}(\bm{0},\bm{\Sigma})$. That is, the transition and observation operators governing the true dynamics are themselves stochastic, so that recovering $(\bm{\mu}^{\star},\bm{u}^{\star})$ from $\bm{u}_{\mathrm{obs}}$ is fundamentally a Bayesian inference problem rather than a deterministic optimisation problem. Therefore, the goal is to find a distribution (usually a two-moment law, most commonly a Gaussian) over the state and parameter vectors that best represents our uncertainty about their true values. In the following subsections, we describe some of the methodologies implemented in Opal to solve this problem, including the classical \ac{kf}, its variants, and variational strategies.

\subsection{Kalman filter}
\label{subs:kf}
The classical \ac{kf} is a sequential algorithm that provides an optimal solution to \eqref{eq: opt problem stochastic} under the assumptions of linearity of both transition and observation operators, and Gaussianity (more on this in the next paragraph). It consists of two main steps: the \textit{forecast} step, where the current state estimate is propagated forward in time using the model dynamics, and the \textit{analysis} step, where the forecast is corrected using new observations. Rather than estimating the state and parameters separately, we consider the augmented variable $\bm{x} \doteq (\bm{\mu},\bm{u})^T \in \R^{N_x}$, where $N_x \doteq N_{\mu} + N_u$, and estimate both jointly with every filter discussed in this and the following subsections. For the time being, we consider an unbiased linear observation model ($\bm{b} \equiv \bm{0}$). Discretising Eq.~\eqref{eq: opt problem stochastic} in time over a partition $t^0 < t^1 < \dots < t^{N_t} \equiv T$ gives
\begin{figure}
  \includegraphics[scale=0.5]{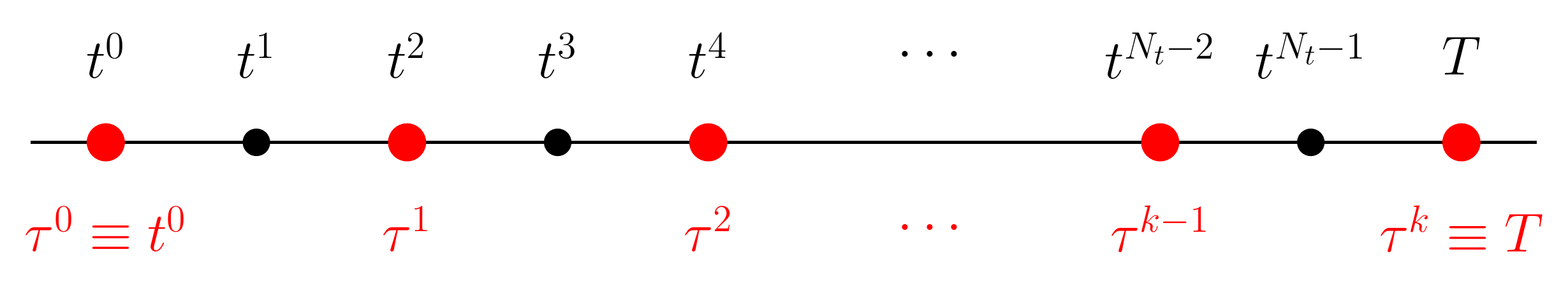}
  \caption{Integration stencil for the transition model (in black) and observation stencil (in red).}
  \label{fig:stencil}
\end{figure}

\begin{align}
  \label{eq: kf forecast model}
  \bm{x}^k &= \bm{\mathcal{F}}_{\bm{\theta}}(\bm{x}^{k-1}), \quad \quad \forall k = 1, \dots, N_t \\
  \label{eq: kf observation model}
  \bm{u}^k_{\mathrm{obs}} &= \bm{H}\bm{x}^k + \bm{\sigma}^k, \ \ \ \quad \forall k = 1, \dots, N_t
\end{align}
\noindent
where $\bm{\mathcal{F}}_{\bm{\theta}}$ is a block-diagonal operator obtained by concatenating (i) a $N_{\mu} \times N_{\mu}$ identity matrix -- acting on the parameter block -- and (ii) the integral of $\bm{\mathcal{A}}_{\bm{\theta}}$ over $(t^{k-1},t^k)$ -- acting on the state variable. $\bm{H} \in \R^{N_{\mathrm{obs}} \times N_x}$ denotes the observation matrix of Subsection~\ref{subs:model}, implicitly augmented with a zero block on the parameter component since observations depend on the state alone. Note that the expressions of both $\bm{\mathcal{F}}_{\bm{\theta}}$ and $\bm{H}$ may depend on the time index $k$, though we will henceforth assume they do not. Also, as shown in Fig.~\ref{fig:stencil}, the observations need not occur at every integration step. On the contrary, it is quite common for observations to be sparse in time ($k < N_t$) and in space ($N_{\mathrm{obs}} < N_u$, and potentially $N_{\mathrm{obs}} \ll N_u$). Sparse observations create challenges, as they increase the complexity of accurately constraining and estimating the evolution of the system \cite{cohn1988observability}. Although we will consider missing observations in the numerical experiments in Section~\ref{sec:examples}, for simplicity, we will assume for the present that observations are available at each integration step. \\

\noindent Consider a single cycle of the recursion, from the previous analysis $\bm{x}^{k-1}_a$ to the next one $\bm{x}^k_a$; to keep the notation light, we drop the time index for the remainder of this derivation. The forecast step propagates the previous analysis through $\bm{\mathcal{F}}_{\bm{\theta}}$, producing a forecast $\bm{x}_f$ and an associated covariance $\bm{P}_f$, which is computed as 
\begin{equation}
  \label{eq: kf forecast covariance}
  \bm{P}_f = \bm{F}\bm{P}_a \bm{F}^{T}+\bm{\Pi},
\end{equation}
where $\bm{F} \in \R^{N_x \times N_x}$ is the Jacobian of $\bm{\mathcal{F}}_{\bm{\theta}}$ at the previous analysis, $\bm{\Pi} \in \R^{N_x \times N_x}$ is the covariance of the process noise (usually block-diagonal, with a zero block on the parameter component), and $\bm{P}_a$ is the previous analysis covariance. We refer to \cite{Kalman1960} for more details. We always assume that the forecast step produces a Gaussian prior\footnote{This is the so-called Gaussian assumption, which assumes (i) Gaussianity of the noise, (ii) Gaussianity of the initial guess, i.e. $\bm{x}_a^0$, and (iii) that the transition operator is ``linear enough'' so that it preserves the Gaussianity of the input.} for $\bm{x}$, that is $\bm{x}\sim\mathcal{N}(\bm{x}_f,\bm{P}_f)$. The analysis step then uses the new observation $\bm{u}_{\mathrm{obs}}$ to characterise the posterior $p(\bm{x}\mid\bm{u}_{\mathrm{obs}})\propto p(\bm{u}_{\mathrm{obs}}\mid\bm{x})\,p(\bm{x})$ via Bayes' theorem. Both the prior and the observation likelihood implied by Eq.~\eqref{eq: kf forecast model} are Gaussian,
\begin{align}
  p(\bm{x}) \propto \exp\left(-\frac12\|\bm{x}-\bm{x}_f\|_{\bm{P}_f^{-1}}^2\right), \qquad
  p(\bm{u}_{\mathrm{obs}}\mid\bm{x}) \propto \exp\left(-\frac12\|\bm{u}_{\mathrm{obs}}-\bm{H}\bm{x}\|_{\bm{\Sigma}^{-1}}^2\right),
\end{align}
\noindent
with the latter being a consequence of the linearity of Eq.~\eqref{eq: kf observation model}. The posterior is itself Gaussian by virtue of Bayes' theorem: 
\begin{align}
  \label{eq: kf map cost}
  p(\bm{x}\mid\bm{u}_{\mathrm{obs}})\propto\exp(-\mathcal{L}_{\mathrm{KF}}(\bm{x})), \qquad 
  \mathcal{L}_{\mathrm{KF}}(\bm{x}) \doteq \frac12\|\bm{x}-\bm{x}_f\|_{\bm{P}_f^{-1}}^2 + \frac12\|\bm{u}_{\mathrm{obs}}-\bm{H}\bm{x}\|_{\bm{\Sigma}^{-1}}^2.
\end{align}
\noindent
The analysis step can be derived by seeking the minimiser of $\mathcal{L}_{\mathrm{KF}}$. Imposing first-order optimality condition and performing some algebraic manipulations (among which the Woodbury identity \cite{GolubVanLoan2013}) gives 
\begin{align}
  \label{eq: kf analysis}
  \bm{x}_a = \bm{x}_f + \bm{K}\left(\bm{u}_{\mathrm{obs}}-\bm{H}\bm{x}_f\right), \qquad
  \bm{P}_a = (\bm{I}-\bm{K}\bm{H})\bm{P}_f, \qquad
  \bm{K} \doteq \bm{P}_f\bm{H}^{T}\left(\bm{H}\bm{P}_f\bm{H}^{T}+\bm{\Sigma}\right)^{-1},
\end{align}
\noindent
where $\bm{K} \in \R^{N_x \times N_{\mathrm{obs}}}$ is the \emph{Kalman gain}, an optimal weighting between the forecast and the (noisy) observation. Equations~\eqref{eq: kf forecast model}--\eqref{eq: kf analysis} define one cycle of the \ac{kf}. Because $\bm{\mathcal{A}}_{\bm{\theta}}$, and hence $\bm{\mathcal{F}}_{\bm{\theta}}$, is generally nonlinear, this linearise-and-correct recursion is more precisely the \ac{ekf} \cite{jazwinski1970stochastic}; it reduces to the classical linear \ac{kf} whenever $\bm{\mathcal{A}}_{\bm{\theta}}$ is itself linear, in which case $\bm{F}$ no longer depends on the point at which it is evaluated. \\

\noindent
The recursion is sequential by construction, updating the joint state-parameter estimate as each new observation arrives, which makes it naturally suited to \ac{da} pipelines. It is, however, subject to several limitations that motivate the variants discussed in the remainder of this section: 
\begin{itemize}
  \item Its optimality (in the minimum-variance sense) only holds under the linear and Gaussian assumptions, which are often violated in practice.
  \item The process- and observation-noise covariances $\bm{\Pi}^k$ and $\bm{\Sigma}^k$ are ordinarily assumed to be constant in time, though they may instead be estimated sequentially when this assumption is untenable \cite{BerrySauer2013}.
  \item Propagating the full covariance $\bm{P}^k_a$ costs $\mathcal{O}(N_x^3)$ per step, which becomes prohibitive for large-scale problems and is the main motivation for the ensemble-based approximations introduced next.
  \item The innovation, $(\bm{u}_{\mathrm{obs}}-\bm{H}\bm{x})$, represents the difference between the observations and their corresponding predictions and is expected, at convergence, to follow a zero-mean Gaussian distribution. This condition cannot be satisfied when the observations are biased; therefore, accounting for such biases requires a non-standard, bias-aware implementation. 
\end{itemize}

\subsection{Ensemble Kalman filter}
\label{subs:enkf}
The \ac{enkf} is a \ac{mc} approximation of the \ac{kf} that uses an ensemble of $N_p$ particles to represent the state distribution. Each particle is propagated through the transition model, and the ensemble statistics are used to compute the moments of the state distribution:
\begin{equation}
  \label{eq:ensemble statistics}
  \bm{x} \approx \widebar{\bm{x}} \doteq \frac{1}{N_p} \sum_{i=1}^{N_p} \bm{x}_i, \quad
  \bm{P} \approx \bm{S} \doteq \frac{1}{N_p-1} \sum_{i=1}^{N_p} (\bm{x}_i - \widebar{\bm{x}})(\bm{x}_i - \widebar{\bm{x}})^T.
\end{equation}
There are several variants of the \ac{enkf}, which differ in how they implement the analysis step. Here, we recall arguably the most frequently used one -- the stochastic \ac{enkf} -- and refer to \cite{SakovOke2008} for the \ac{denkf} and to \cite{evensen2009data} for the square-root \ac{enkf}. Once again dropping the time index for simplicity, the analysis of the stochastic \ac{enkf} reads as follows:
\begin{align}
  \label{eq: enkf analysis}
  \bm{x}^{(i)}_a = \bm{x}^{(i)}_f + \bm{K}\left(\bm{u}^{(i)}_{\mathrm{obs}}-\bm{H}\bm{x}^{(i)}_f\right), \qquad 
  \bm{u}^{(i)}_{\mathrm{obs}} \doteq \bm{u}_{\mathrm{obs}} + \bm{\sigma}^{(i)},
  \qquad i = 1,\hdots,N_p,
\end{align}
where $\bm{\sigma}^{(i)}$ is a realisation of the observation noise, i.e. $\bm{\sigma}^{(i)} \sim \mathcal{N}(\bm{0},\bm{\Sigma})$. This perturbation of the observation is a key feature of the stochastic \ac{enkf}, and aims to avoid the underestimation of the analysis covariance \cite{burgers1998analysis}, a common issue in ensemble-based methods. The Kalman gain $\bm{K}$ is computed as in \eqref{eq: kf analysis}, with $\bm{P}_f\bm{H}^T$ and $\bm{H}\bm{P}_f\bm{H}^T$ replaced by their ensemble counterparts:
\begin{equation}
  \bm{P}_f\bm{H}^{T} \equiv \frac{1}{N_p-1} \sum_{i=1}^{N_p} (\bm{x}^{(i)}_f - \widebar{\bm{x}}_f)(\bm{y}^{(i)} - \widebar{\bm{y}})^T, \qquad
  \bm{H}\bm{P}_f\bm{H}^{T} \equiv \frac{1}{N_p-1} \sum_{i=1}^{N_p} (\bm{y}^{(i)} - \widebar{\bm{y}})(\bm{y}^{(i)} - \widebar{\bm{y}})^T,
\end{equation}
where $\bm{y}^{(i)} \doteq \bm{H}\bm{x}^{(i)}_f$ is the predicted observation for the $i$-th particle. Crucially, $\bm{P}_f$ is never formed explicitly, so the analysis costs $\mathcal{O}(N_p N_x N_{\mathrm{obs}})$ per cycle rather than $\mathcal{O}(N_x^3)$, a significant saving whenever $N_p \ll N_x$. The \ac{enkf} is also asymptotically consistent: in the limit $N_p \to \infty$, its solution converges to the optimal one, even for nonlinear dynamics \cite{evensen2009data}. In practice, $N_p$ is kept far below $N_x$, and accuracy at small ensemble sizes is recovered through inflation \cite{sun2024high,anderson2009spatially}, localisation \cite{sun2024high,bishop2017gain}, and adaptivity \cite{berry2013adaptive}, which are discussed in the next subsection.

\subsection{Auxiliary EnKF methodologies: inflation, localisation, and adaptivity}
\label{subs:infl_loc_adapt}
A well-known pathology of the \ac{enkf} is that $\bm{S}$, being a finite-sample estimate of $\bm{P}$ built from only $N_p \ll N_x$ members, systematically underestimates the true forecast spread, which may cause the ensemble to collapse around a (potentially biased) estimate and stop responding to observations. Multiplicative covariance inflation \cite{anderson2009spatially} is the simplest remedy: before the analysis step, every forecast deviation from the ensemble mean is rescaled by a factor $\rho > 1$,
\begin{equation}
  \label{eq: multiplicative inflation}
  \bm{x}_f^{(i)} \leftarrow \widebar{\bm{x}}_f + \rho\left(\bm{x}_f^{(i)} - \widebar{\bm{x}}_f\right), \qquad i = 1,\hdots,N_p,
\end{equation}
which leaves the ensemble mean unchanged while inflating the sample covariance by $\rho^2$, i.e. $\bm{S}_f \leftarrow \rho^2\bm{S}_f$. More advanced methods in the literature consider time-varying inflation factors, for example the one developed in \cite{sun2024high}. Here, the authors estimate a new value of $\rho$ at every cycle, together with a localisation of the ensemble covariance, directly from the data. Localisation techniques have been proposed to dampen spurious long-range correlations that are sometimes visible in $\bm{S}_f$ due to the finite ensemble size \cite{hamill2001distance}. Usually, they replace $\bm{S}_f$ by a localised version $\bm{\mathcal{L}}\circ\bm{S}_f$, where $\bm{\mathcal{L}}$ is a distance-based correlation-tapering operator (e.g. of Gaspari--Cohn type \cite{GaspariCohn1999}). In Opal, we implement the following iterative updates, proposed in \cite{sun2024high}:
\begin{enumerate}
  \item Estimate $\bm{\mathcal{L}}$ by choosing a length scale such that the resulting localised covariance is consistent with the original $\bm{S}_f$.
  \item Estimate $\rho$ as the \ac{mle} of an inflated, localised version of \eqref{eq: enkf analysis}, which amounts to minimising
  \begin{equation}
    \label{eq: nll inflation localisation}
    \mathrm{NLL}(\rho) \doteq \frac12\bm{d}^{T}\bm{C}(\rho)^{-1}\bm{d} + \frac12\log\det\bm{C}(\rho), \qquad
    \bm{C}(\rho) \doteq \rho^2\bm{H}\left(\bm{\mathcal{L}}\circ\bm{S}_f\right)\bm{H}^{T}+\bm{\Sigma}.
  \end{equation}
\end{enumerate}
\noindent
Finally, \cite{berry2013adaptive} addresses a complementary problem: rather than explicitly adjusting the ensemble spread, optimal values for the noise covariances $\bm{\Pi}$ and $\bm{\Sigma}$ are adaptively sought to recover the optimality of the filter, as detailed in \cite{Mehra1970}. The authors have likened this approach to a form of \emph{additive} inflation, as opposed to the multiplicative inflation discussed earlier. In practice, this lets $\bm{\Pi}$ and $\bm{\Sigma}$ track genuine changes in model or instrument error over the course of an experiment, rather than remaining fixed at whatever values were assumed at initialisation -- which is the default procedure in standard \acp{kf}.

\subsection{Bias-aware Kalman filters}
\label{subs:bias-aware}
Bias-aware \acp{kf} are a class of methods that explicitly account for systematic errors in the filtering procedure, usually by including a bias term in the observation space. A sizeable portion of Opal is dedicated to bias-aware implementations, largely following the approaches in \cite{novoa2022realtime,Novoa2023}\footnote{We note that these references discuss the specific case of a bias-aware \ac{enkf}. However, as we believe the principles can be extended to other filter variants, the present subsection employs a broader perspective than our references.}. In particular, the latter reference employs a bias-aware observation operator of the form \eqref{eq: bias-aware obs}, and seeks to minimise
\begin{align}
  \label{eq: bias-aware enkf map cost}
  \mathcal{L}_{\mathrm{BA-KF}}(\bm{x}) \doteq \mathcal{L}_{\mathrm{KF}}(\bm{x}) + \frac{\gamma}{2}\|\bm{b}\|_{\bm{\Xi}^{-1}}^2,
\end{align}
where $\gamma \geq 0$ is a regularisation parameter and $\bm{\Xi} \in \R^{N_{\mathrm{obs}} \times N_{\mathrm{obs}}}$ is a covariance matrix for the bias, commonly chosen as $\bm{\Xi} \equiv \bm{\Sigma}$. The analysis step and the expression of the Kalman gain are then modified to minimise \eqref{eq: bias-aware enkf map cost}: 
\begin{equation}
  \label{eq: bias-aware kf analysis}
  \begin{split}
    \bm{x}_a &= \bm{x}_f + \bm{K}\left((\bm{I} + \bm{J})(\bm{u}_{\mathrm{obs}}-\bm{H}\bm{x}_f - \bm{b}) -\gamma \bm{\Sigma} \bm{\Xi}^{-1}\bm{J}\bm{b}\right), \\
    \bm{K} &\doteq \bm{P}_f\bm{H}^{T}\left( \left((\bm{I} + \bm{J})(\bm{I} + \bm{J})^T + \gamma \bm{\Sigma} \bm{\Xi}^{-1}\bm{J}\bm{J}^T\right) \bm{H}\bm{P}_f\bm{H}^{T}+\bm{\Sigma}\right)^{-1},
  \end{split}
\end{equation}
where $\bm{J} \doteq \frac{\partial \bm{b}}{\partial (\bm{H}\bm{x}_f)}$. Specifically, in \cite{Novoa2023}, the authors model the bias with an \ac{esn} \cite{jaeger2001echo}, which is a recurrent neural network with a sparsely connected hidden layer with random entries. In \cite{Novoa2023}, the authors argue for the great suitability of \acp{esn} in the context of bias modelling for \ac{kf} pipelines, given (i) their ability to capture time-evolving processes \cite{novoa2022realtime}, (ii) their relatively low computational cost (also for training), and (iii) the straightforward computation of $\bm{J}$. We recall that a standard \ac{esn} at the $k$th time step may be formulated as 
\begin{equation}
  \label{eq:esn}
  \bm{b}^k(\bm{y}) = \bm{W}_{\mathrm{out}}\bm{r}^k(\bm{y}), \quad 
  \bm{r}^k(\bm{y}) = (1-\alpha)\bm{r}^{k-1}(\bm{y}) + \alpha f( \bm{W}_{\mathrm{in}}\bm{y}^k + \bm{W}_{\mathrm{hid}}\bm{r}^{k-1}), \qquad 
  \bm{r}^k \in \R^{N_r}, \ \bm{y}^k \in \R^{N_{\mathrm{obs}}}, 
\end{equation}
where $\alpha \in [0,1]$, $f$ is the activation function (usually a hyperbolic tangent), the matrices $\bm{W}_{\mathrm{in}} \in \R^{N_r\times N_{\mathrm{obs}}}$ and $\bm{W}_{\mathrm{hid}} \in \R^{N_r \times N_r}$ are two fixed matrices, $\bm{W}_{\mathrm{out}} \in \R^{N_{\mathrm{obs}} \times N_r}$ is the only trainable parameter of the network, $\bm{r}$ is the so-called hidden state, $\bm{y}$ is the input to the network, and $\bm{b}$ is clearly the output. For the precise formulation of the \ac{esn} employed throughout this work, we refer to \cite{Novoa2023}, which considers a more complex formulation than that in Eq.~\eqref{eq:esn}. There are two possible configurations for the \ac{esn}: 
\begin{itemize}
  \item \textit{open-loop}: the input variable is set to a given (training) value. This configuration is used during the training phase.
  \item \textit{closed-loop}: the input variable is set to the previous output of the network, that is $\bm{y}^k \doteq \bm{b}^{k-1}$. This configuration is used whenever we wish to evaluate the \ac{esn} -- i.e. the bias model -- once it is trained.
\end{itemize}
The reason why training an \ac{esn} is inexpensive is that it reduces to a ridge regression problem \cite{lukosevicius2012practical}
\begin{equation}
  \label{eq:esn training}
  \bm{W}_{\mathrm{out}} = \arg\min_{\bm{W}} \sum_{k=1}^{N_{\mathrm{train}}} \left\| \bm{b}^k - \bm{W}\bm{r}^k(\bm{y}) \right\|_2^2 + \lambda\|\bm{W}\|_2^2 \ \Longrightarrow \ \bm{W}_{\mathrm{out}} = \bm{B}\bm{R}^{T}(\bm{R}\bm{R}^{T} + \lambda\bm{I})^{-1},
\end{equation}
which is far more efficient than backpropagation with gradient descent methods. In Eq.~\eqref{eq:esn training}, $\lambda \geq 0$ represents the Tikhonov regularisation parameter, whereas $\bm{B} \in \R^{N_{\mathrm{obs}} \times N_{\mathrm{train}}}$ is a matrix of training biases (obtained by concatenating $N_{\mathrm{train}}$ observed biases during the training phase) and $\bm{R} \in \R^{N_r \times N_{\mathrm{train}}}$ is a matrix of training hidden states (obtained by running \eqref{eq:esn} in closed-loop on the observed biases). \\ 
Although several considerations have been omitted in the present work for brevity, the overall procedure for constructing and running a bias-aware \ac{kf} is summarised in Alg.~\ref{alg:bias-aware filter}. For a detailed explanation, we refer to the original work \cite{Novoa2023}.
\begin{algorithm}[t]
  \centering
  \caption{Bias-aware Kalman filter}\label{alg:bias-aware filter}
  \begin{algorithmic}[1]
    \State Initialise training data: $\left(\bm{R},\bm{B}\right) = \left([],[]\right)$ \Comment{ESN training}
    \For{$k = 1:N_{\mathrm{train}}$}
    \State Run transition: $\left(\bm{x}^k,\bm{b}^k\right) = \left(\bm{\mathcal{F}}_{\bm{\theta}}(\bm{x}^{k-1}), \bm{u}_{\mathrm{obs}}^k - \bm{H}\bm{x}^k\right)$
    \State Accumulate training data: $\left(\bm{R},\bm{B}\right) \leftarrow \left([\bm{R}, \bm{r}^k(\bm{b}^k)],[\bm{B}, \bm{b}^k]\right)$
    \EndFor
    \State Train \ac{esn} with ridge regression \eqref{eq:esn training}
    \For{$k = N_{\mathrm{train}} + 1 : N_{\mathrm{train}} + N_{\mathrm{wash}}$}  \Comment{ESN washout}
    \State Run transition: $\left(\bm{x}^k,\bm{b}^k\right) = \left(\bm{\mathcal{F}}_{\bm{\theta}}(\bm{x}^{k-1}), \bm{u}_{\mathrm{obs}}^k - \bm{H}\bm{x}^k\right)$
    \State Update hidden state \eqref{eq:esn}: $\bm{r}^k = \bm{r}^k(\bm{b}^{k-1})$
    \EndFor
    \For{$k = N_{\mathrm{train}} + N_{\mathrm{wash}} + 1 : N_{\mathrm{train}} + N_{\mathrm{wash}} + N_t$} \Comment{Bias-aware DA}
    \State Run \ac{esn} in closed-loop: $\left(\bm{r}^k,\bm{b}^k\right) = \left(\bm{r}^k(\bm{b}^{k-1}), \bm{W}_{\mathrm{out}}\bm{r}^k(\bm{b}^{k-1})\right)$
    \State Compute $\bm{J}$ as outlined in \cite{Novoa2023}
    \State Run transition: $\bm{x}_f^k = \bm{\mathcal{F}}_{\bm{\theta}}(\bm{x}_a^{k-1})$
    \State Run analysis \eqref{eq: bias-aware kf analysis}
    \EndFor
  \end{algorithmic}
\end{algorithm}

\subsection{Incorporating reduced error models into the Kalman filter}
\label{subs:rem}
When the transition model $\bm{\mathcal{A}}_{\bm{\theta}}$ is replaced by a computationally cheaper surrogate -- e.g. a \ac{rom} \cite{quarteroni2014reduced,negri2015reduced,doi:10.1137/22M1509114,MUELLER2024115767,MUELLER2026116790} -- to make repeated evaluations within the \ac{da} pipeline tractable, an appropriate filtering technique is needed to account for the resulting surrogate error. While this could in principle be handled by the bias-aware filter of Subsection~\ref{subs:bias-aware}, the \ac{rom} error is more naturally treated as a zero-mean, parameter-dependent quantity that is small near the training samples and grows predictably away from them. To this end, we adopt the approach of \cite{pagani2017efficient}, though we note that this is more a working choice than a definitive strategy, since identifying the most suitable strategy for surrogate-error-aware \ac{da} remains an open question we intend to keep exploring. However, the latter already provides a practical, computationally inexpensive framework for the problems considered here. \\

\noindent
The approach proposed in \cite{pagani2017efficient} can be briefly summarised as follows:
\begin{enumerate}
  \item A dictionary of \ac{fe} solutions $\{\bm{u}^k(\bm{\mu}_i)\}_{i}^{N_{\mathrm{train}}}$ -- with $k = 1,\hdots,N_t$ -- is generated at a set of training parameters $\{\bm{\mu}_i\}_{i=1}^{N_{\mathrm{train}}}$. Of these, a number $N_{\mathrm{off}} \leq N_{\mathrm{train}}$ of offline parameters are used to construct an $n$-dimensional \ac{rom} subspace $\mathrm{span}\{\bm{\phi}_i\}_{i=1}^n \subset \R^{N_u}$, with $n \ll N_u$, such that
  \begin{equation}
    \label{eq:rom}
    \frac{d\widehat{\bm{u}}(\bm{\mu})}{dt} + \widehat{\bm{\mathcal{A}}}(\widehat{\bm{u}}(\bm{\mu});\bm{\mu}) = \bm{0}, \ \text{ where hopefully } \ \bm{u}_n(\bm{\mu}) \doteq \sum_{i=1}^n \widehat{u}_i(\bm{\mu})\bm{\phi}_i \approx \bm{u}(\bm{\mu}) \ \ \forall \bm{\mu} \in \{\bm{\mu}_i\}_{i=1}^{N_{\mathrm{off}}}.
  \end{equation}
  For a complete review of the \ac{rom} methodologies used to derive Eq.~\eqref{eq:rom}, we refer to the ``classics'' \cite{quarteroni2014reduced,negri2015reduced}, as well as to our works \cite{doi:10.1137/22M1509114,MUELLER2024115767,MUELLER2026116790}, and to \cite{muellerbadiagridaproms}, which describes the implementation details of GridapROMs.jl, our in-house \ac{rom} Julia library that we exploit for the practical computation of \acp{rom}.
  \item We consider a biased observation operator as in \eqref{eq: bias-aware obs}, with the bias term being the \ac{rom} error, i.e. 
  \begin{equation}
    \bm{b} \equiv \bm{b}(\bm{\mu}) \doteq \bm{G}\left(\bm{u}(\bm{\mu}) - \bm{u}_n(\bm{\mu})\right).
  \end{equation}
  In practice, as the value of $\bm{\mu}$ is unknown, $\bm{b}$ needs to be efficiently estimated across the entire parameter space. To this end, \cite{pagani2017efficient} consider the \ac{blup} associated with the random process $\{\bm{b}(\bm{\mu}_i)\}_{i=1}^{N_{\mathrm{cal}}}$, with $N_{\mathrm{cal}} \leq N_{\mathrm{train}}$ being the number of calibration parameters, obtained via kriging-based interpolation \cite{chiles1999geostatistics}. In essence, the \ac{blup} -- here referred to as $\bm{\beta}$ -- is computed as a linear combination of the bias at the calibration parameters:
  \begin{equation}
    \bm{\beta}(\bm{\mu}) \doteq \sum_{i=1}^{N_{\mathrm{cal}}} w_i(\bm{\mu}) \bm{b}(\bm{\mu}_i) ,
  \end{equation}
  where $\{w_i\}_{i=1}^{N_{\mathrm{cal}}}$ are the kriging weights found by solving a constrained optimisation problem -- minimisation of the variance of $\bm{\beta} - \bm{b}$, while imposing $\bm{\beta}$ to be zero-mean -- over the set of calibration parameters \cite{pagani2017efficient}. 
\end{enumerate}
The analysis step and the expression of the Kalman gain of a standard \ac{kf} are modified as follows: 
\begin{equation}
  \label{eq: rom-bias-aware kf analysis}
  \bm{x}_a = \bm{x}_f + \bm{K}\left(\bm{u}_{\mathrm{obs}}-\bm{H}\bm{x}_f - \bm{\beta}\right), \qquad
  \bm{K} \doteq \bm{P}_f\bm{H}^{T}\left( \bm{H}\bm{P}_f\bm{H}^{T}+\bm{\Sigma}+\bm{\Gamma}\right)^{-1},
\end{equation} 
where $\bm{\Gamma}$ is the matrix of trace variances associated with $\bm{\beta}$ \cite{pagani2017efficient}. Overall, this approach is easily incorporated within a \ac{kf} pipeline, since it only requires minor modifications to Eq.~\eqref{eq: rom-bias-aware kf analysis}. The approach is computationally inexpensive, since the \ac{rom} training and the kriging weights computation are performed during the offline phase, assuming that $N_{\mathrm{cal}} \approx N_{\mathrm{off}}$, so that the number of required snapshots $N_{\mathrm{train}}$ is limited. We also note that rigorous \emph{a priori} error bounds for the \ac{rom} have been used in the literature \cite{manzoni2016accurate} for the same purpose, but strong assumptions on the formulation of the \ac{pde} are required. Another approach that we may consider in the future is to estimate the bias term $\bm{b}$ directly from the \ac{rom} residual.

\subsection{Other Bayesian methods: unscented Kalman filter and particle filters}
\label{subs:misc}
For completeness, we close this section with a brief overview of the remaining methodologies implemented in Opal. So far, we have mentioned the \ac{ekf} (linearisation of operators) and \ac{enkf} (\ac{mc} approximation) as ways to handle potential nonlinearity in the system. The \ac{ukf} \cite{julier1997new,wan2000unscented} uses a different approach to achieve the same goal: it propagates a small, deterministic set of \emph{sigma points} through the nonlinear map $\bm{\mathcal{F}}_{\bm{\theta}}$, and reconstructs $\bm{x}_f$ and $\bm{P}_f$ from the transformed points. This captures the effect of nonlinearity on the first two moments to a higher order than the \ac{ekf} linearisation, at a comparable cost, while the analysis step itself remains unchanged from Eq.~\eqref{eq: kf analysis}. \\

\noindent
Despite sharing many similarities with \acp{kf} (particularly the \ac{enkf}), particle filters dispense with the Gaussian assumption entirely, representing $p(\bm{x}\mid\bm{u}_{\mathrm{obs}})$ by a weighted sample $\{\bm{x}^{(i)},w^{(i)}\}_{i=1}^{N_p}$ rather than by its first two moments. Weights are updated multiplicatively with the observation likelihood, $w^{(i)} \propto w^{(i)}\,p(\bm{u}_{\mathrm{obs}}\mid\bm{x}_f^{(i)})$, and the particles are periodically resampled to counteract \emph{weight degeneracy}, the well-known tendency of all but a few weights to collapse to zero. Opal implements \ac{sir} and the regularised particle filter discussed in \cite{BerrySauer2013}. We finally recall that no linearity assumption is placed on $\bm{\mathcal{A}}_{\bm{\theta}}$ or $\bm{\mathcal{G}}_{\bm{\sigma}}$, and that particle filters are asymptotically exact as $N_p \to \infty$.

\subsection{Variational methods}
\label{subs:variational}
Variational \ac{da} methods recast the state-parameter estimation problem as a gradient-based optimisation procedure. In essence, a trajectory $\{\bm{x}^k\}_k$ is sought to minimise, over a given time window, a cost function measuring the discrepancy between:
\begin{itemize}
  \item The available observations and the predicted observations obtained by applying the observation operator to the trajectory, i.e. $\bm{\mathcal{H}}(\bm{x}^k)$.
  \item The trajectory's initial state and a background estimate, which represents our prior knowledge about the system.
\end{itemize}
We tackle two common formulations. First, the 4D-Var \cite{talagrand1987variational} seeks an initial condition whose model trajectory provides the best fit to observations distributed over the entire assimilation window $(t^0,T)$, that is
\begin{equation}
  \label{eq: variational cost 4dvar}
  \mathcal{L}_{\mathrm{4d-var}}(\bm{x}^0) \doteq \frac12\|\bm{x}^0-\bm{x}_b\|_{\bm{B}^{-1}}^2 + \frac12\sum_{k=1}^{N_t}\|\bm{u}^k_{\mathrm{obs}} - \bm{\mathcal{H}} \left(\bm{x}^k\right)\|_{\bm{\Sigma}^{-1}}^2, \ \text{ where } \ \bm{x}^k = \bm{\mathcal{F}}(\bm{x}^{k-1}),
\end{equation}
where $\bm{x}_b \in \R^{N_x}$ is the aforementioned background estimate, and $\bm{B} \in \R^{N_x \times N_x}$ is its associated covariance. Second, the 3D-Var \cite{lorenc1986analysis} iteratively seeks an estimate of the state at one single time step at the time:
\begin{equation}
  \label{eq: variational cost 3dvar}
  \mathcal{L}_{\mathrm{3d-var}}(\bm{x}^0) \doteq \frac12\|\bm{x}^0-\bm{x}_b\|_{\bm{B}^{-1}}^2 + \frac12\|\bm{u}^k_{\mathrm{obs}} - \bm{\mathcal{H}} \left(\bm{x}^k\right)\|_{\bm{\Sigma}^{-1}}^2, \ \text{ where } \ \bm{x}^k = \bm{\mathcal{F}}(\bm{x}^{k-1}), \quad k = 1, \ldots, N_t.
\end{equation}
After each optimisation step, the value of the background estimate is updated to the solution obtained at the previous time step, that is $\bm{x}_b \leftarrow \bm{x}^{k-1}$ for every $k \geq 2$. The minimisation of Eqs.~\eqref{eq: variational cost 4dvar}-\eqref{eq: variational cost 3dvar} is performed very efficiently in Opal, by exploiting the automatic differentiation capabilities of the packages SciMLSensitivity \cite{SciMLSensitivity} and GridapTopOpt.jl \cite{GridapTopOpt} for \acp{ode} and \acp{pde}, respectively, to compute the loss gradients. The ensuing gradient descent-like minimisation procedure is then performed with the BFGS algorithm \cite{broyden1970convergence}. 
\begin{remark}
  Though not apparent from Eq.~\eqref{eq: variational cost 4dvar}, the 4D-Var in general still operates sequentially as most \ac{da} strategies. Indeed, once new observations become available past $T$, the 4D-Var is re-run over a new assimilation window, with the initial condition set to the previously estimated $\bm{x}^{N_t}$ from \eqref{eq: variational cost 4dvar}. 
\end{remark}
\section{Implementation details}
\label{sec:implementation}
This section describes how the methodologies of Subsection~\ref{sec:description} are realised in code. Subsection~\ref{subs:types} introduces the main abstractions around which the whole package is organised. Subsection~\ref{subs:sciml-gridap} details the native interoperability layer with the SciML and Gridap/GridapROMs ecosystems, through which an arbitrary ODE or transient, full- or reduced-order PDE solution can serve as a transition model with no change to the surrounding assimilation code. Subsection~\ref{subs:high-level} presents the high-level API used to assemble a complete \ac{da} experiment, and Subsection~\ref{subs:rc} describes the self-contained, cache-efficient \ac{rc} implementation underlying the bias-aware filter of Subsection~\ref{subs:bias-aware}. The section closes, in Subsection~\ref{subs:usage-example}, with a complete usage example. \\ 
Since most implementation efforts in Opal are devoted to Bayesian filtering techniques, rather than variational methods, we focus most of our attention on the former in this section. In any case, the latter are implemented in a similar fashion, and the interested reader is referred to the source code for further details.

\subsection{Types and hierarchy}
\begin{figure}[t]
\centering
\resizebox{\textwidth}{!}{%
\begin{forest}
[Law\{D\}, abs
  [FirstMoment\\{\tiny$\equiv$Law\{1\}}, alias
    [GenericFirstMoment, conc]
    [Particle, conc]
  ]
  [SecondMoment\\{\tiny$\equiv$Law\{2\}}, alias
    [NormalLaw, conc]
    [UniformLaw, conc]
    [SigmaPoints, conc]
    [Ensemble, conc]
  ]
  [ConstrainedLaw\{D\}, wrap]
]
\end{forest}}
\caption{\myfont{Law} hierarchy. In red: abstract types; in orange: type aliases; in purple: parametric subtypes; and in blue: concrete subtypes.}
\label{fig:tree-law}
\end{figure}
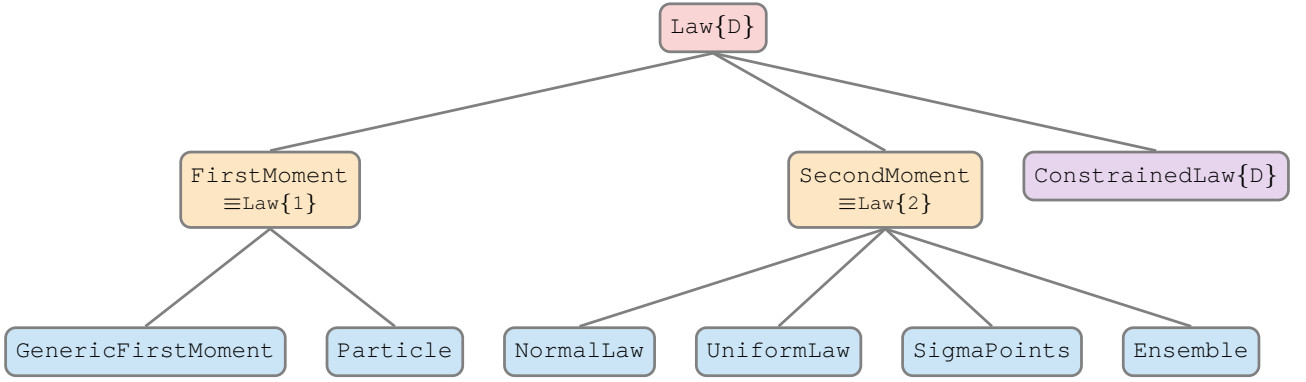

\begin{figure}[t]
\centering
\resizebox{\textwidth}{!}{%
\begin{forest}
[Model\{A\}, abs
  [LinearModel\\{\tiny$\equiv$Model\{Linear\}}, alias
    [AlgebraicModel, conc]
    [TrivialLinearModel, abs
      [ZeroModel, conc]
      [IdentityModel, conc]
    ]
  ]
  [NonlinearModel\\{\tiny$\equiv$Model\{Nonlinear\}}, alias
    [GenericModel, conc]
    [TaperModel, conc]
    [DifferentialModel, abs
      [ODEModel\\{\tiny SciML}, conc]
      [TransientPDEModel\\{\tiny Gridap/GridapROMs}, conc]
    ]
  ]
  [MemoryModel\{A\}, wrap]
]
\end{forest}}
\caption{\myfont{Model} hierarchy. In red: abstract types; in orange: type aliases; in purple: parametric subtypes; and in blue: concrete subtypes.}
\label{fig:tree-model}
\end{figure}
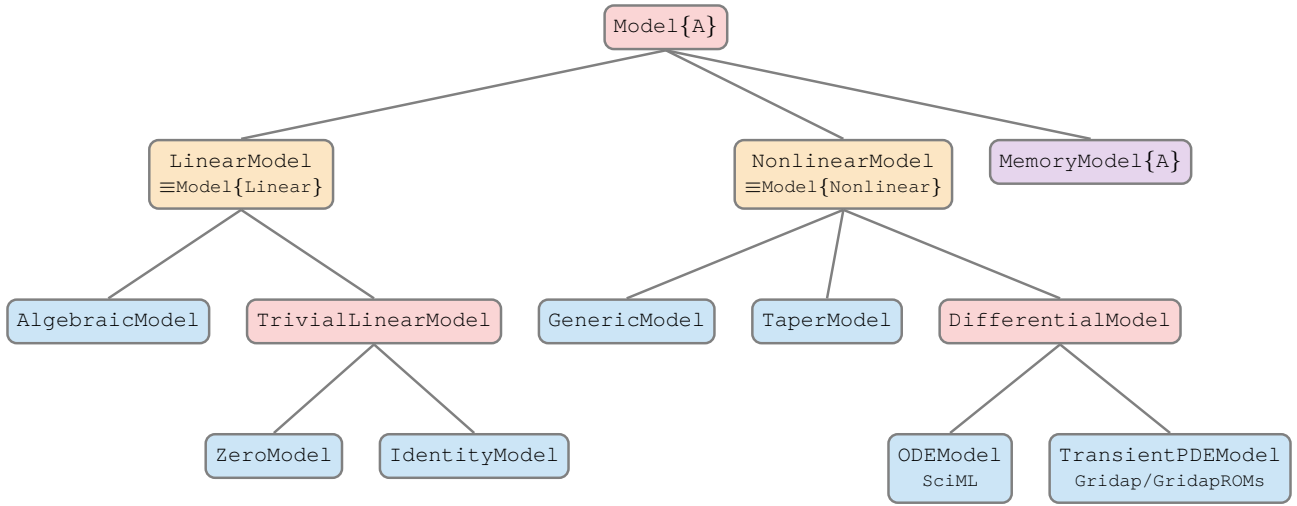

A considerable portion of Opal is organised around a small number of abstract types, summarised in \tab{tb: abstract types meteo}. A \myfont{Law} is an interface for probability distributions of parameters and states that can be efficiently propagated through a \myfont{Model}, which in general represents any map, and is mostly used to describe a transition or observation operator. The \myfont{Law} hierarchy is shown in \fig{fig:tree-law}. The leaves of the tree (i.e. concrete types) are the distributions used throughout the methods described in Section~\ref{sec:description}: a \myfont{NormalLaw} is the default type for the prior propagated throughout a \ac{kf} and for the representation of noise distributions, \myfont{Ensemble} is an ensemble law used for \acp{enkf}, and similarly \myfont{SigmaPoints}, \myfont{Particle} and \myfont{GenericFirstMoment} are respectively used for \acp{ukf}, particle filters, and variational methods. The last distribution we mention is the \myfont{ConstrainedLaw}, which wraps a \myfont{Law} and a set of constraints that are used to enforce bounds on the state variables after propagation. This is useful, for instance, when the unknown parameters are known to belong to a certain parameter space. \\ 
\label{subs:types}
\begin{center}
	\small
	\begin{table}[t]
		\begin{tabular}{lll}
			\toprule
			&\multicolumn{1}{l}{Abstract type}
			&\multicolumn{1}{l}{Purpose}
			\\
			\midrule
      &\myfont{Law}
			&Probabilistic representation of state/parameter variables
			\\
			&\myfont{Model}
			&Forward/observation operators (e.g. custom linear/nonlinear maps, SciML \acp{ode}, Gridap \acp{pde})
			\\
			&\myfont{DAMethod}
			&Abstraction shared by \ac{da} methods
			\\
			\bottomrule
		\end{tabular}
		\caption{Main abstract types involved in Opal.}
		\label{tb: abstract types meteo}
	\end{table}
\end{center}
The \myfont{Model} hierarchy in \fig{fig:tree-model} shows how these types are categorised depending on whether they represent a linear map (\myfont{LinearModel}) or a nonlinear one (\myfont{NonlinearModel}). For the former, the most important example is the \myfont{AlgebraicModel}, which simply wraps the matrix; for the latter, we mention (i) the \myfont{TaperModel}, which represents the tapering operator $\bm{\mathcal{L}}$ discussed in Subsection~\ref{subs:infl_loc_adapt}, (ii) the \myfont{ODEModel} and \myfont{TransientPDEModel} used whenever the transition is defined by \acp{ode} and transient \acp{pde}, and (iii) the \myfont{GenericModel}, which wraps a generic Julia function (or a Gridap \myfont{Map} \cite{Badia2020}). Another notable subtype is the \myfont{MemoryModel}, which wraps a \myfont{Model} and a \myfont{Law}; we will clarify the usefulness of the latter in Section~\ref{sec:examples}.  \\ 
The user may think of a \myfont{Model} as an efficient \myfont{Law}-to-\myfont{Law} map (although it also works as an \myfont{AbstractArray}-to-\myfont{AbstractArray} mapping). The efficiency is driven by a cache-friendly design, which allows the user to preallocate any buffer needed for its evaluation and then reuse it in every subsequent call. This clearly separates the \myfont{Model} from a plain Julia function, and is a crucial design choice in the context of \ac{da}, where the transition and observation operators are evaluated repeatedly in a loop, often over an ensemble of states. The system of caches is implemented following a similar design as in Gridap, namely by extending the \myfont{return\_cache} and \myfont{evaluate!} methods implemented therein. \\

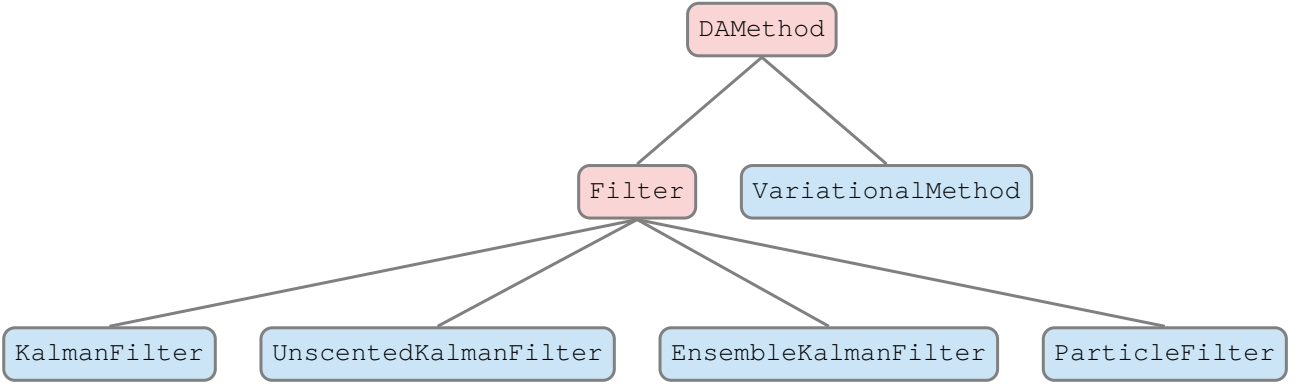
\begin{figure}[t]
\centering
\resizebox{\textwidth}{!}{%
\begin{forest}
[DAMethod, abs
  [Filter, abs
    [KalmanFilter, conc]
    [UnscentedKalmanFilter, conc]
    [EnsembleKalmanFilter, conc]
    [ParticleFilter, conc]
  ]
  [VariationalMethod, conc
  ]
]
\end{forest}}
\caption{\myfont{DAMethod} hierarchy. In red: abstract types; and in blue: concrete subtypes.}
\label{fig:tree-filter}
\end{figure}

\begin{figure}[t]
	\includegraphics{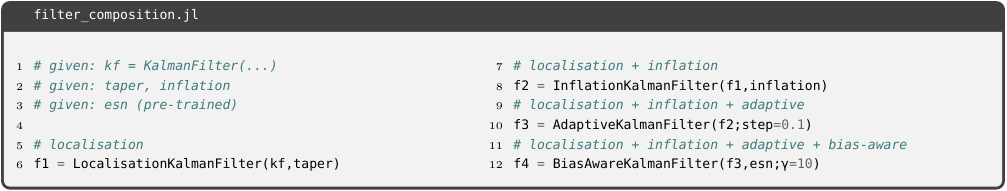}
	\caption{Examples of composite \myfont{Filter}s in Opal.}
	\label{lst: filter composition}
\end{figure}

\noindent
Every one of the methods described throughout Section~\ref{sec:description} is implemented either as a \myfont{Filter} or as a \myfont{VariationalMethod}, both subtypes of a common \myfont{DAMethod} abstraction. Conceptually, a \myfont{DAMethod} is an object defined by a (transition, observation) \myfont{Model} pair, a (prior, observation prior) \myfont{Law} pair, a (process noise, observation noise) \myfont{Law} pair, a set of methods that implement the forecast and analysis steps, and lastly a cache structure collecting all the necessary buffers to perform these steps in-place. Given a \myfont{DAMethod} and some observations, the \ac{da} can be launched simply by calling the \myfont{loop} method, whose basic implementation is shown in \lst{lst: filter loop}. 
\begin{figure}[t]
	\includegraphics{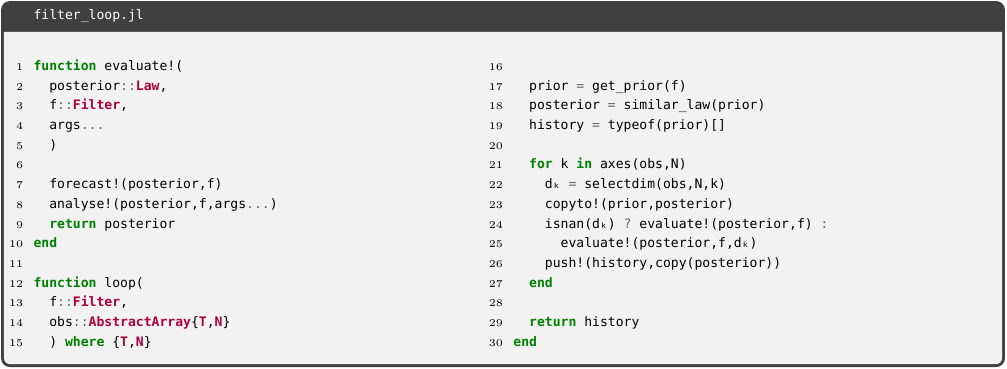}
	\caption{Simplified implementation of the \ac{da} loop for \myfont{Filter} subtypes. Though a similar implementation holds for \myfont{VariationalMethod}, the latter dispenses with the forecast-then-analyse paradigm.}
	\label{lst: filter loop}
\end{figure} 
As the name suggests, the latter implements a \emph{for} loop over the observations (assumed to be stacked along their last axis), whereupon the posterior is continuously updated according to the \myfont{forecast!} and \myfont{analyse!} methods specified for \myfont{Filter}s, and via an optimisation on sliding windows for \myfont{VariationalMethod}s. A vector of posterior \myfont{Law}s -- referred to throughout this work as the filter's history -- is the output of \myfont{loop}. The roles of \myfont{forecast!} and \myfont{analyse!} can be summarised in \tab{tb: analysis methods}.
\begin{center}
  \small
  \begin{table}[t]
    \begin{tabular}{lllll}
      \toprule
      &\multicolumn{1}{l}{Method}
      &\multicolumn{1}{l}{Parent method}
      &\multicolumn{1}{l}{Purpose}
      &\multicolumn{1}{l}{Output}
      \\
      \midrule
      &\myfont{transition!}
      &\myfont{forecast!}
      &Runs the transition \myfont{Model} on the posterior from the previous time step
      &$\bm{x}_f^k = \bm{\mathcal{F}}_{\bm{\theta}}(\bm{x}_a^{k-1})$
      \\
      \midrule
      &\myfont{observation!}
      &\myfont{analyse!}
      &Runs the observation \myfont{Model} on the output of \myfont{forecast!}
      &$\bm{H}\bm{x}_f^k$
      \\
      &\myfont{innovation!}
      &\myfont{analyse!}
      &Computes the innovation by removing the current observation
      &$\bm{u}_{\mathrm{obs}}-\bm{H}\bm{x}_f^k$
      \\
      &\myfont{kalman\_gain!}
      &\myfont{analyse!}
      &Computes the Kalman gain
      &$\bm{K}$
      \\
      &\myfont{update!}
      &\myfont{analyse!}
      &Computes the analysed posterior \myfont{Law}
      &$\bm{x}_a^k$
      \\
      \bottomrule
    \end{tabular}
    \caption{Workflow of \myfont{forecast!} and \myfont{analyse!}, the two core methods in the \myfont{DAMethod} class.}
    \label{tb: analysis methods}
  \end{table}
\end{center}
The \myfont{DAMethod} hierarchy is summarised in \fig{fig:tree-filter}. The leaves of the tree are the concrete implementations of some of the methods described in Section~\ref{sec:description}: \myfont{KalmanFilter}, \myfont{UnscentedKalmanFilter}, \myfont{EnsembleKalmanFilter}, and \myfont{ParticleFilter} are respectively the implementations of the \ac{kf}, \ac{ukf}, \ac{enkf}, and particle filter; while \myfont{VariationalMethod} constitutes the backbone for our variational methods\footnote{We note that 3D-Var and 4D-Var share the same exact structure up to the definition of their respective assimilation windows (see Eqs.~\eqref{eq: variational cost 4dvar}-\eqref{eq: variational cost 3dvar}), and thus we do not implement them separately.}. The \ac{ekf} is essentially a \myfont{KalmanFilter} -- which represents the standard \ac{kf} as presented in Subsection~\ref{subs:kf} -- whose transition and/or observation operators are subtypes of \myfont{NonlinearModel}. The remaining methodologies previously discussed -- namely localisation, inflation, adaptivity, bias-awareness and \ac{rom} calibration -- are implemented as composable wrappers around a \myfont{Filter} instance. These wrappers are themselves \myfont{Filter} subtypes that hold an inner \myfont{Filter}, and therefore only need to specialise a small set of methods (usually, just the ones in \tab{tb: analysis methods}). The seamless filter composition and relatively simple type hierarchies allow for a very flexible and modular design, with minimal effort required to implement new methods or combine existing ones. Examples of \myfont{Filter} composition are shown in \fig{lst: filter composition}.

\subsection{Native integration with the SciML and Gridap ecosystems}
\label{subs:sciml-gridap}
In this subsection, we focus on the specific cases where the transition operator is an \ac{ode} or a transient \ac{pde}, which comprise a large portion of the applications of interest in the inverse modelling and \ac{da} communities. We begin with the former. Arguably the most commonly used package for \acp{ode} in the Julia ecosystem is SciML's OrdinaryDiffEq.jl \cite{OrdinaryDiffEq}, which implements a wide range of time-stepping algorithms and provides a common interface for their use. The \myfont{ODEWrapper} implemented in Opal is a thin wrapper around any OrdinaryDiffEq.jl integrator, and the \myfont{Model} constructor dispatches on it to produce a \myfont{ODEModel} that can be used as a transition operator in any \myfont{DAMethod} described in the previous subsection. We show in \lst{lst: sciml integration} the code implementing this design, along with a minimum-working example constructing a \ac{kf}. The crucial line of code is the $6$th of the listing, and the remaining ones are verbatim. A complete usage example involving \ac{ode}-based transition operators is discussed in Subsection~\ref{subs:usage-example}. During the \myfont{loop} function, the transition operator is advanced using the Tsitouras $5/4$ Runge-Kutta time integrator \cite{TSITOURAS2011770} implemented in OrdinaryDiffEq.jl. \\ 

\noindent
As shown in \lst{lst: gridap integration}, a very similar design principle also holds for transient \acp{pde}. In this case, the general constructor \myfont{Model} (line $9$) dispatches to \myfont{TransientPDEModel} when provided with an \myfont{ODESolution} input (defined at line $8$), a Gridap type. Essentially everything works in the same manner as for the \ac{ode} case, with the only difference being that \myfont{forecast!} propagates the \myfont{Law} in time through a \ac{fe} integrator, assembler, and solver rather than through an \ac{ode} counterpart. We note that the syntax of \lst{lst: gridap integration} also allows an \myfont{ODESolution} object to wrap a \ac{rom} generated by GridapROMs. This makes it possible to run a computationally cheaper version of the transition operator at each iteration of \myfont{loop}, while preserving exactly the same interface and syntax as in \lst{lst: gridap integration}.

\begin{figure}[t]
	\includegraphics{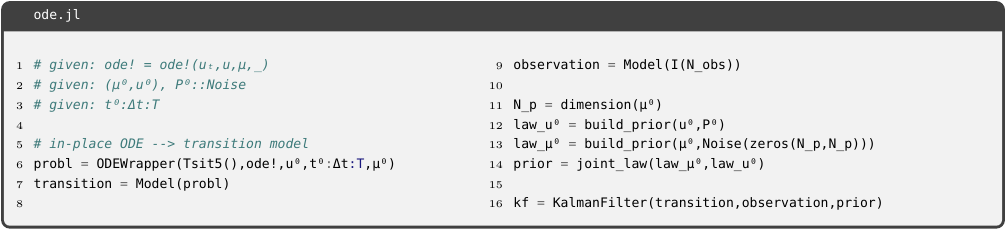}
	\caption{Defining a \myfont{Filter} for an \ac{ode}-constrained problem in Opal.}
	\label{lst: sciml integration}
\end{figure} 	

\begin{figure}[t]
	\includegraphics{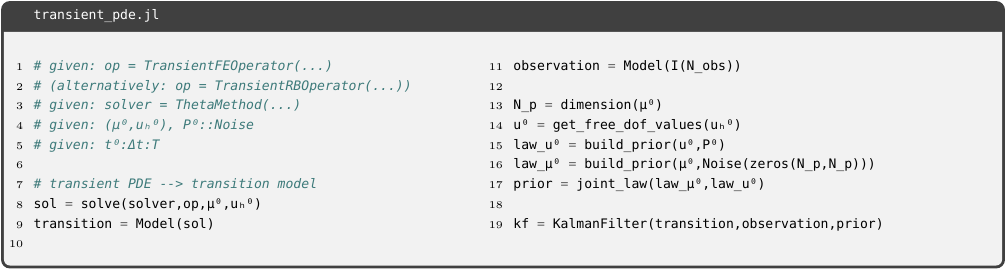}
	\caption{Defining a \myfont{Filter} for a transient \ac{pde}-constrained problem in Opal.}
	\label{lst: gridap integration}
\end{figure}

\subsection{High-level interface}
\label{subs:high-level}
Opal is designed to give the user a simple yet elegant interface for assembling a complete \ac{da} experiment without engaging directly with the machinery of Subsection~\ref{subs:types}. This is not a cosmetic convenience: standard experiments typically involve several distinct time intervals -- a spin-up window, an assimilation window and, for the bias-aware filter of Subsection~\ref{subs:bias-aware}, a training, a washout, and a ``spread'' window \cite{Novoa2023} as well -- each possibly sampled at a different rate than the observations. We encapsulate this bookkeeping in two structures, \myfont{TimeStencils} and \myfont{StencilArray}, and expose the resulting workflow through a handful of routines, such as \myfont{execute}, \myfont{warmup!}, \myfont{build\_prior}, and \myfont{build\_observations}, shown in \lst{lst: high-level interface}. 
\begin{figure}[t]
	\includegraphics{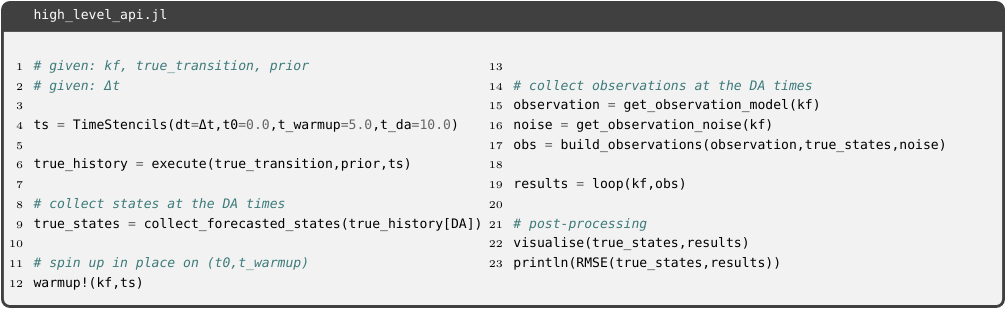}
	\caption{High-level interface in Opal. An instance of \myfont{TimeStencils} may be used to define a \myfont{StencilArray} representing the history of a transition operator (\myfont{true\_history}, line $6$), which can be lazily indexed as in line $9$ and propagated forward in time (line $12$). The API also allows for easy construction of observations, observation operators, and running \ac{da}.}
	\label{lst: high-level interface}
\end{figure} 
A \myfont{TimeStencils} object partitions an experiment into up to six contiguous, named phases -- \myfont{WARMUP}, \myfont{TRAIN}, \myfont{WASHOUT}, \myfont{SPREAD}, and \myfont{DA}, plus \myfont{ALL} for the entire window -- built by chaining user-supplied durations (\texttt{t\_warmup}, \texttt{t\_train}, \texttt{t\_wash}, \texttt{t\_spread}, \texttt{t\_da}) one after another from a starting time \texttt{t0}. Each phase is stored as a pair of endpoints and materialised into a concrete time grid on demand: indexing \myfont{ts[WARMUP]}, or \myfont{ts[TRAIN:SPREAD]} for a contiguous span of phases, returns the corresponding range at the model time step \texttt{dt}. Every phase constant has an \myfont{OBS}-prefixed counterpart (\myfont{OBSWARMUP}, \ldots, \myfont{OBSALL}) denoting the very same interval sampled at the observation time step \texttt{dt\_obs} instead -- the two time scales illustrated in \fig{fig:stencil} -- so that a single \myfont{TimeStencils} instance simultaneously drives an integrator at its natural time step and aligns observations to their own, generally coarser, grid. \\
Rather than having every routine repeat this bookkeeping, \myfont{execute} and its relatives return their result wrapped in a \myfont{StencilArray}, a lazy container tagging an array with the \myfont{TimeStencils} instance and the phase over which it was computed. Re-indexing a \myfont{StencilArray} into a different phase does not recompute the trajectory: it calls \myfont{restrict} or \myfont{expand}, depending on whether the requested phase's grid is coarser or finer than the one currently stored, to resample the time axis of the underlying array onto the new grid. This is what allows, for instance, a full-resolution \myfont{WARMUP} trajectory and its coarser \myfont{OBSWARMUP} sub-sample to be views of a single computation rather than two separate runs. \\
More specifically, \myfont{execute} runs a transition \myfont{Model} forward from a prior \myfont{Law} and returns the resulting \myfont{StencilArray}; \myfont{warmup!} advances a \myfont{DAMethod}'s own prior over a phase in-place, discarding the trajectory, which is the idiomatic way to spin up an \ac{ode}/\ac{pde} integrator (an analogous syntax may be employed for a \myfont{EchoStateNetwork}, a type discussed in the next subsection); \myfont{build\_prior} constructs the initial \myfont{Law} appropriate to the target filter from a seed state and covariance; and \myfont{build\_observations} applies an observation \myfont{Model} to a state history to produce synthetic or real observations, optionally corrupted by noise or an additive bias term. Once a \myfont{DAMethod} and an observation array are available, \myfont{loop} runs the assimilation and returns a \myfont{DAResults} object, on which \myfont{visualise} and a small library of diagnostics -- RMSE, normalised RMSE, negative log-likelihood, the NEES/NIS consistency scores, spread-skill ratio, innovation autocorrelation, and rank histograms -- operate directly.

\begin{figure}[t]
	\centering
	\subfloat[Accuracy]{\includegraphics[width=0.49\textwidth]{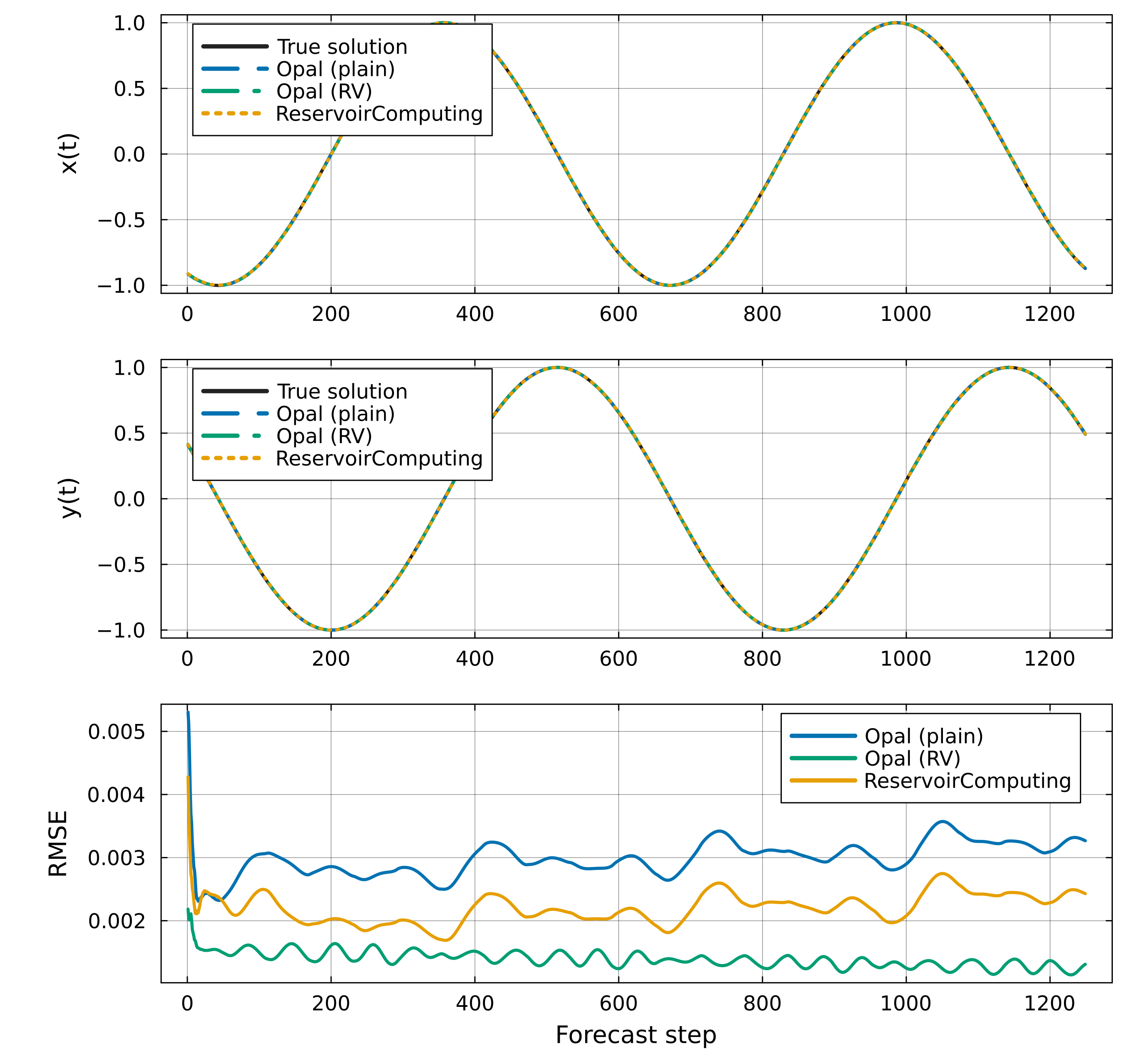}}
	\hfill
	\subfloat[Efficiency]{\includegraphics[width=0.49\textwidth]{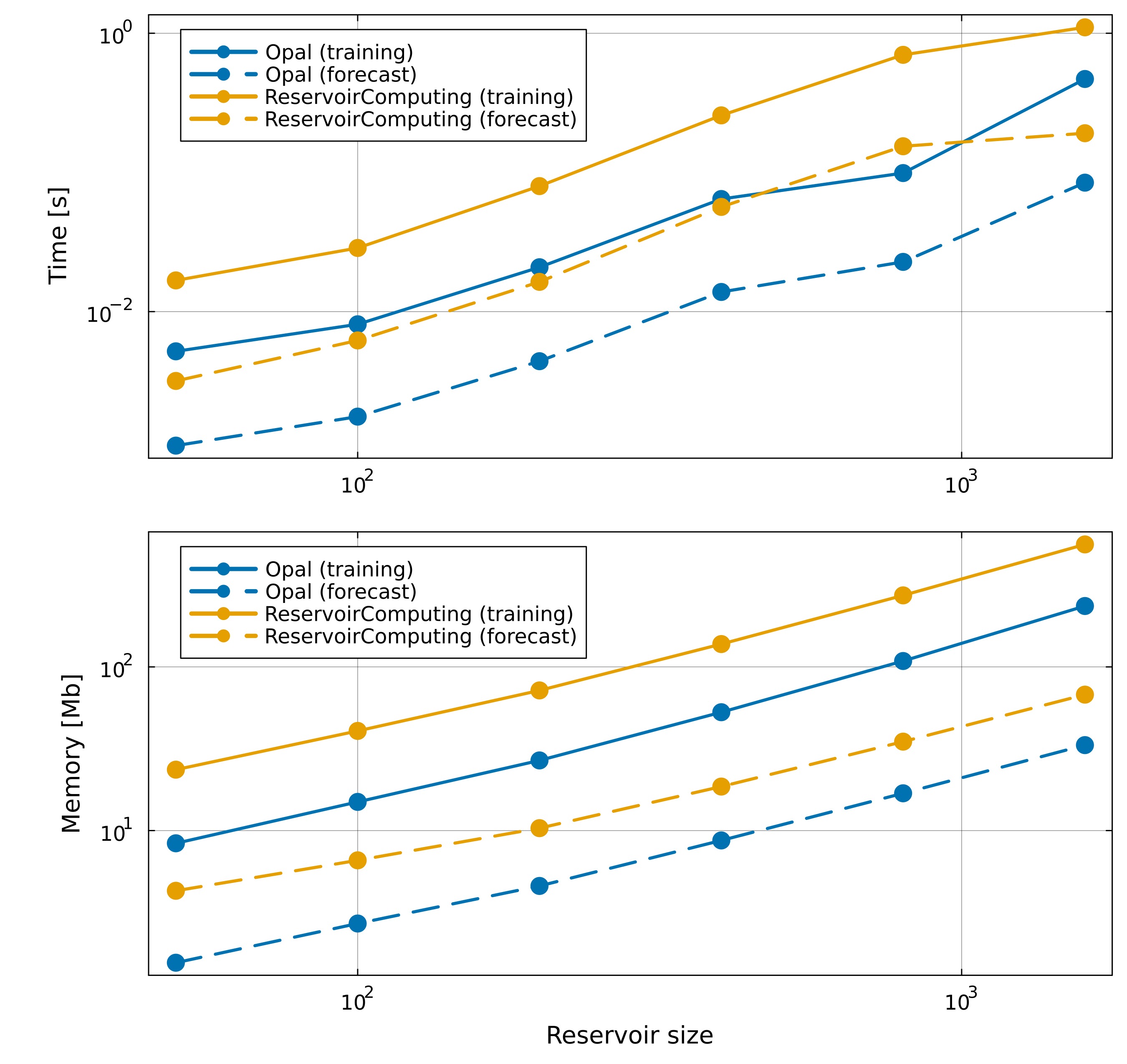}}
  \caption{Comparison of Opal's \ac{esn} implementation with ReservoirComputing.jl on the Hopf Bifurcation system \cite{guckenheimer1983nonlinear}. (A) Accuracy comparison in terms of the predicted $x(t)$ and $y(t)$ state coordinates against the true solution, and in terms of RMSE. (B) Efficiency (in terms of time and memory) comparison based on the time required to train the \ac{esn} on a training set as a function of the number of hidden units. The results show that Opal's implementation is both more accurate (when using the recycle-validation (RV) strategy) and more efficient than ReservoirComputing.jl. To ensure a fair efficiency comparison, we consider only the plain Opal implementation, since RV training is inherently more computationally expensive, while the forecasting stage is identical in both Opal settings.}
	\label{fig: rc comparison}
\end{figure}

\subsection{Efficient reservoir computing}
\label{subs:rc}
The \ac{esn} bias model described in Subsection~\ref{subs:bias-aware} is implemented from scratch, rather than by relying on ReservoirComputing.jl \cite{Martinuzzi2022} -- arguably the most widely used \ac{rc} library in the Julia ecosystem. Opal does reuse two of its utility routines for the initialisation of the fixed weights $\bm{W}_{\mathrm{in}}$ and $\bm{W}_{\mathrm{hid}}$ \eqref{eq:esn}, but the forward pass, the ridge-regression training \eqref{eq:esn training}, and, more generally, the type hierarchy are a self-contained implementation built around the same caching paradigm discussed in Subsection~\ref{subs:types} for \myfont{Model} subtypes. This improves the efficiency of the models implemented in ReservoirComputing.jl, as shown in Fig.~\ref{fig: rc comparison}. Additionally, a major contribution with respect to ReservoirComputing.jl is the possibility of training the \ac{esn} with the \emph{recycle-validation} strategy proposed in \cite{Novoa2023} for the training of a more accurate bias model. We stress that these are architectural choices motivated by the specific repeated-evaluation setting of the previously discussed bias-aware filtering, and our implementation does not represent a general-purpose substitute for ReservoirComputing.jl, which supports a broader range of \ac{rc} architectures and training strategies. 

\subsection{Usage example}
\label{subs:usage-example}
\begin{figure}[t]
	\includegraphics{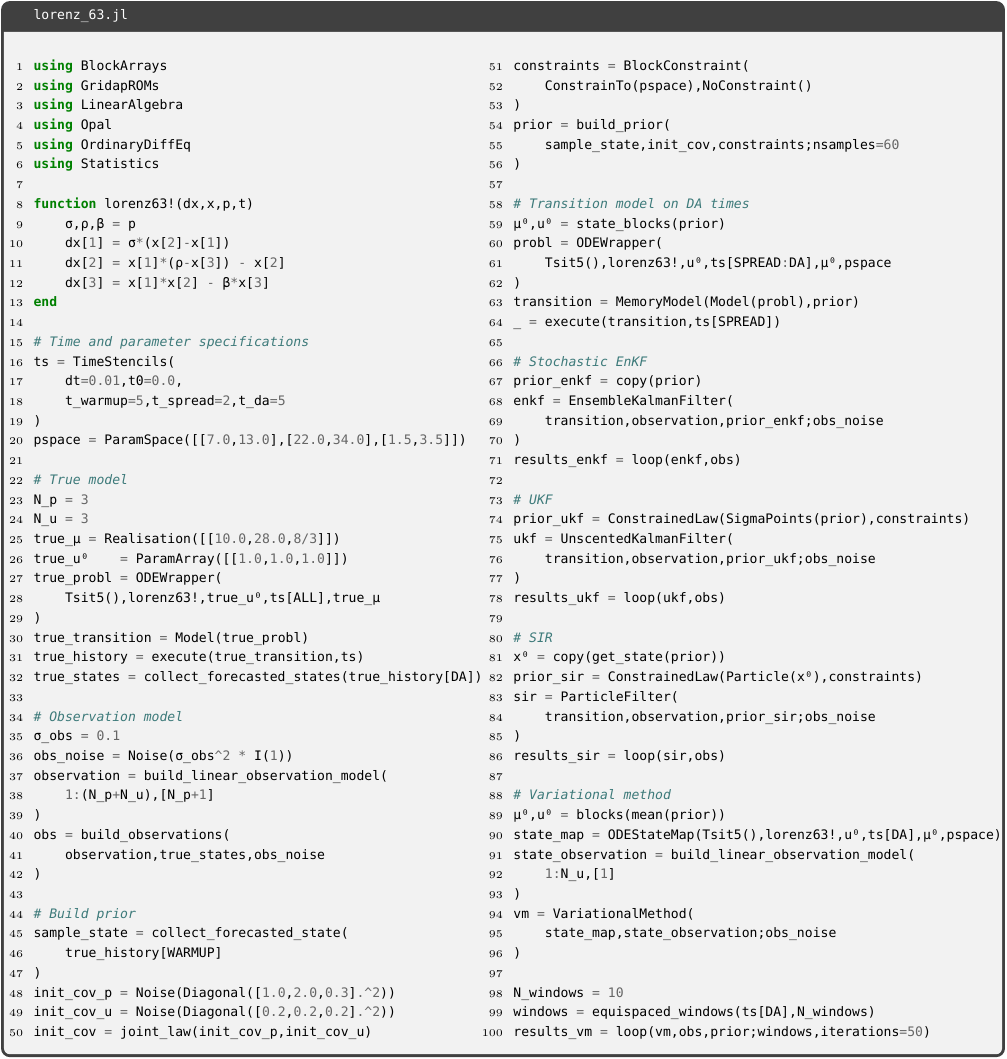}
	\caption{Solving the Lorenz-63 \ac{da} problem with Opal.}
	\label{lst: usage example}
\end{figure}

In this usage example, we solve a joint parameter-state estimation problem based on the Lorenz-63 equations (lines $8-13$) using four \ac{da} strategies: (i) the stochastic \ac{enkf}, (ii) the \ac{ukf}, (iii) the \ac{sir} particle filter, and (iv) a variational method. The system jointly estimates three parameters alongside three state variables, and the ``true'' system is assumed to be the classical chaotic regime obtained for $\bm{\mu} \doteq (10,28,8/3)^T$. Let us address point-by-point the implementation highlights of the script:
\begin{itemize}
  \item A synthetic trajectory is generated by integrating the true model forward in time\footnote{The \myfont{Realisation} and \myfont{ParamArray} types used for the parameter and initial condition are native to GridapROMs.} (line $31$) from an initial condition equal to $(1,1,1)^T$. At line $32$, the states are extracted from the trajectory in the \myfont{DA} interval using the high-level function \myfont{collect\_forecasted\_states}.
  \item The observation operator (defined at lines $37-39$ using \myfont{build\_linear\_observation\_operator}) maps the joint six-dimensional parameter-state vector to its fourth entry, i.e. the first state variable $u_1$; observations are corrupted by Gaussian noise and collected (lines $40-42$).
  \item An ensemble prior (lines $44-56$) is derived by adding Gaussian noise to the ensemble state at the end of a warmup phase, with parameters constrained to prescribed ranges via \myfont{BlockConstraint} (lines $51-53$), whereas the state variables are unconstrained. The \myfont{joint\_law} function -- called at line $50$ -- concatenates the two independent noise distributions for $\bm{\mu}$ and $\bm{u}$, thus forming a joint distribution.
  \item The transition model (lines $58-64$) is defined as a \myfont{MemoryModel} object, initially mentioned in Subsection~\ref{subs:types}. This type is reserved for models which can be idiomatically propagated in time before the start of the \ac{da} window, e.g. because they require a warmup phase to be correctly initialised. In this case, the model is propagated over the \myfont{SPREAD} window, which is a common practice in order to add sufficient diversity to the ensemble members \cite{SakovOke2008}. The same syntax is used to define the true transition model at line $30$, which is used to generate the synthetic observations.
  \item Lastly, the four \ac{da} methods -- \ac{enkf} (lines $66-71$), \ac{ukf} (lines $73-78$), \ac{sir} (lines $80-86$), and variational (lines $88-100$) -- are defined and executed. The three Bayesian filters share essentially the same implementation, up to the definition of a different prior \myfont{Law} and the choice of the \myfont{Filter} subtype. The variational method differs only in that it does not pass a standard transition model, rather it passes an \myfont{ODEStateMap} which represents the $\bm{\mu} \to \bm{u}$ mapping -- needed both for time integration and also to efficiently compute the derivatives of the loss functions \eqref{eq: variational cost 4dvar}-\eqref{eq: variational cost 3dvar} with respect to $\bm{\mu}$ -- and an observation operator defined for the non-augmented state. To run the \myfont{loop} for the variational method, in addition to the usual observations, we also need to pass the sliding windows (lines $98-99$) and the prior, which represents the background term previously called $\bm{x}_b$ in Eqs.~\eqref{eq: variational cost 4dvar}-\eqref{eq: variational cost 3dvar}.
\end{itemize}
The snippet demonstrates Opal's high-level interface, which allows the user to assemble a complete parameter-state \ac{da} experiment -- including four different algorithms -- with very few lines of expressive code. We underline the central role of \myfont{TimeStencils} in order to achieve this. We will present the numerical results of this experiment in Subsection~\ref{subs:lorenz63}.
\section{Numerical examples}
\label{sec:examples}
\noindent
In this section, we present four numerical examples that demonstrate the ability of Opal to tackle complex \ac{da} problems with a high-level interface while achieving good accuracy. In Subsection~\ref{subs:lorenz63}, we report the numerical results obtained from the Lorenz-63 model already discussed in Subsection~\ref{subs:usage-example}. In Subsection~\ref{subs:van der pol}, we construct a \myfont{BiasAwareFilter} to solve a Van der Pol oscillator problem with biased observations. In Subsection~\ref{subs:navier stokes}, we demonstrate the discovery of the Reynolds number of a turbulent Navier-Stokes flow around a square cavity. Finally, in Subsection~\ref{subs:heat equation}, we showcase \ac{da} with model surrogate techniques for the Heat equation, and compare the results obtained with and without the error calibration discussed in Subsection~\ref{subs:rem}. Although we aim to provide the complete code snippets for each test case, some lines that are not essential to understanding the examples are omitted for brevity. The omitted scripts are provided in the appendix, and readers are referred to the original source code for the complete implementations.

\subsection{Reporting the numerical results from the Lorenz 63 model}
\label{subs:lorenz63}

\begin{figure}[t]
  \centering
  {\legenddot{black}~True\quad\legenddot{plotblue}~\ac{enkf}\quad\legenddot{plotred}~\ac{ukf}\quad\legenddot{plotgreen}~\ac{sir}\quad\legenddot{plotpurple}~Variational}\par\smallskip
  \includegraphics[scale=0.27]{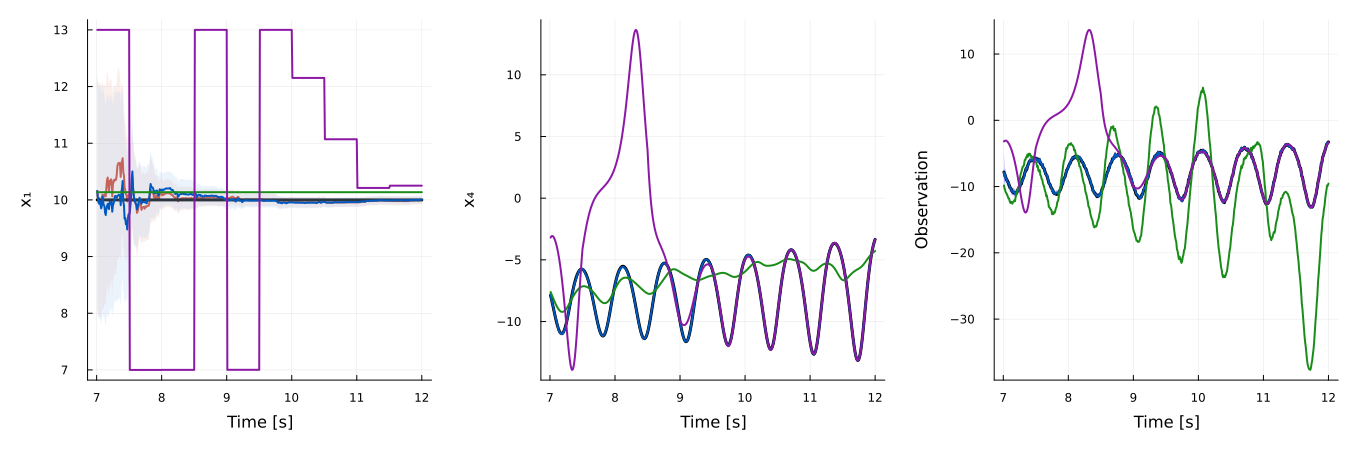}
  \caption{Lorenz-63 example. From left to right: (i) estimation of $x_1 \equiv \mu_1$, (ii) estimation of $x_4 \equiv u_1$, and (iii) estimation of observations. We note that, in Opal, the parameter estimate from the variational method is piece-wise constant across each assimilation window.}
  \label{fig:lorenz-63}
\end{figure}

We briefly dedicate this part to the discussion of the results -- presented in Fig.~\ref{fig:lorenz-63} -- obtained when executing the script in \lst{lst: usage example}. Here, we compare the performance of the four \ac{da} methods in the estimation of the first parameter and first state variable ($x_1$ and $x_4$ respectively), as well as the comparison in the observation space. The true values are shown alongside the \ac{da} predictions, with a confidence interval of $2\sigma$ for the \ac{enkf} and \ac{ukf} (visible only in the first plot, whereas the interval collapses in the remaining ones), since these two propagate a \myfont{SecondMoment} throughout the loop; on the other hand, the \ac{sir} and variational methods do not propagate any uncertainty, and therefore no confidence interval is shown. The \ac{enkf} and \ac{ukf} are able to accurately estimate the true unknowns, both in the augmented state space and in the observation space. The same is true for the variational method, although about $4$ assimilation windows are needed for the estimates to converge to the true values for $x_4$ and the observations, and longer still for $x_1$. Another downside of the latter method is that its simulations have cost us $\sim \mathcal{O}(10^3)$ more in time than the Bayesian filters, even with the automatic differentiation capabilities we exploit from SciMLSensitivity. On the other hand, the \ac{sir} is unable to accurately estimate all three unknowns, as its predictions diverge from the true values after a few time steps. We believe that \ac{sir}, and more generally our \myfont{ParticleFilter}s, are less effective than the \ac{kf} variants in parameter-state estimation problems, and they perform better in state-only estimation problems -- as demonstrated by the flat estimate of $x_1$. In addition, the application of \ac{sir} here is less natural than the remaining \ac{da} methods, since the Gaussian assumption holds here. \\
The better performance of \ac{enkf} and \ac{ukf}, as well as the nature of our numerical tests, motivate our choice of using the \ac{enkf} as the main \ac{da} strategy we employ in the next subsections.

\subsection{Bias-aware EnKF on the Van der Pol oscillator}
\label{subs:van der pol}

\begin{figure}[t]
  \centering
  \includegraphics{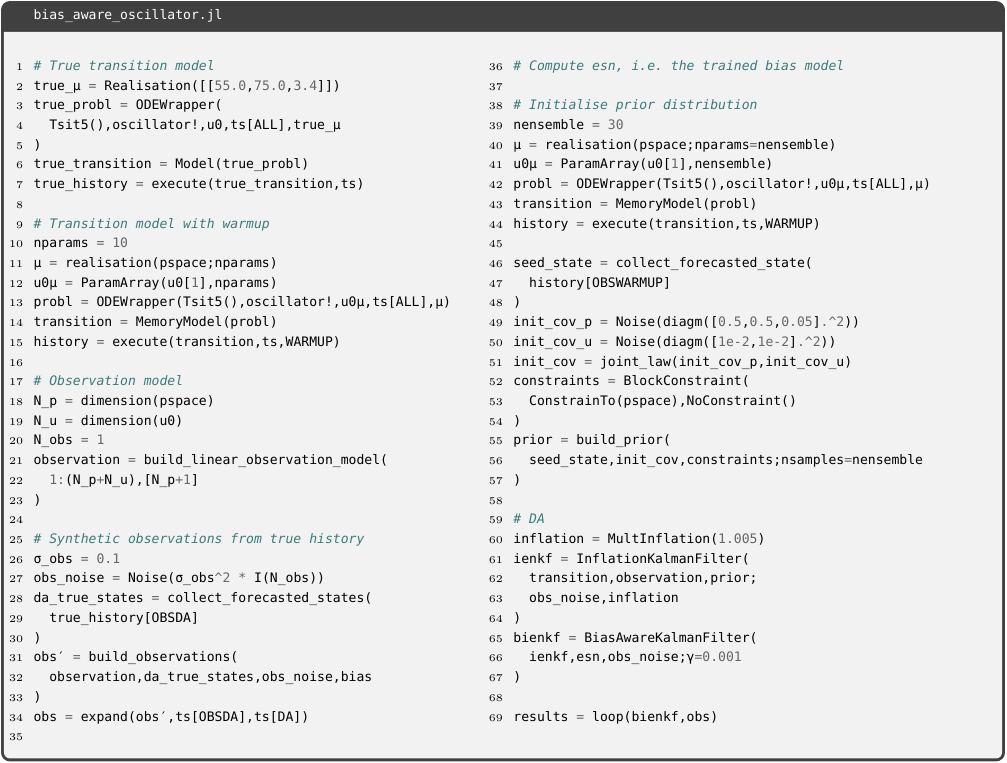}
  \caption{Code snippet used to run the \ac{da} for the Van der Pol example.}
  \label{lst:van der pol}
\end{figure}

\begin{figure}[t]
  \centering
  \includegraphics{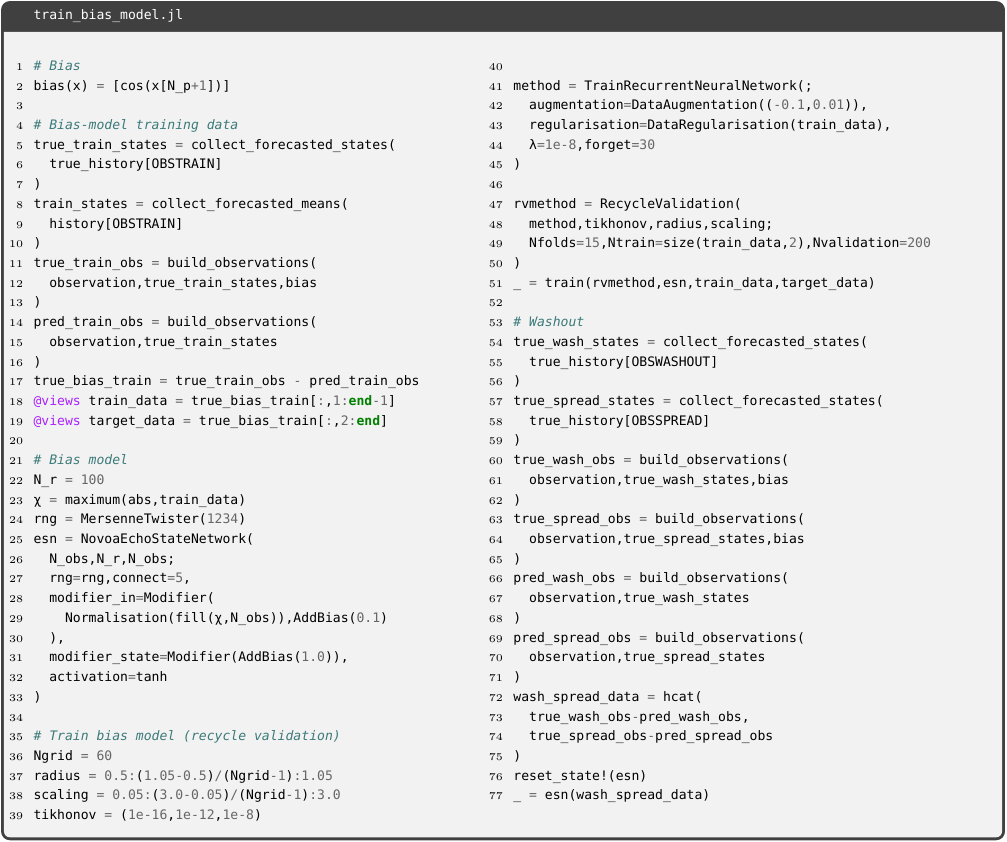}
  \caption{Code snippet used to define and train the bias model for the Van der Pol example.}
  \label{lst:van der pol bias}
\end{figure}

\begin{figure}[t]
  \centering
  {\legenddot{black}~True\quad\legenddot{plotred}~Bias-unaware \ac{enkf}\quad\legenddot{plotblue}~Bias-aware \ac{enkf}}\par\smallskip
  \includegraphics[scale=0.27]{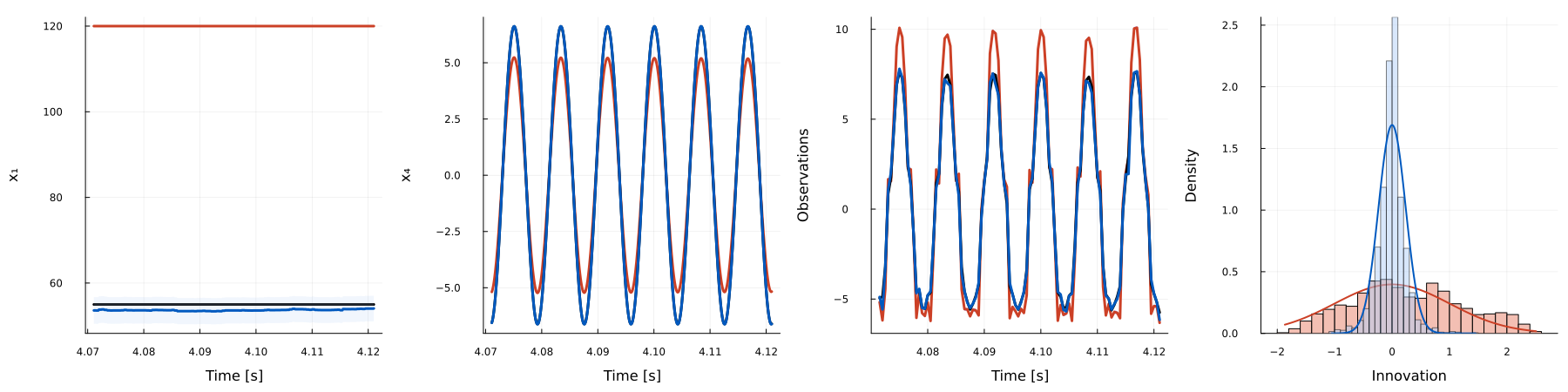}
  \caption{Van der Pol example. From left to right: (i) estimation of $x_1 \equiv \mu_1$ in the last $0.5s$ of the assimilation window, (ii) estimation of $x_4 \equiv u_1$ in the last $0.5s$ of the assimilation window, (iii) estimation of observations in the last $0.5s$ of the assimilation window, and (iv) estimated probability density function of the innovation.}
  \label{fig:van der pol}
\end{figure}

In this example, we consider a bias-aware \ac{enkf} procedure for the state-parameter estimation of a Van der Pol oscillator \cite{vanderPol1926} with biased observations. This numerical test is inspired by \cite{Novoa2023}, which employs the same approach to a similar benchmark. The transition and observation models are given by:
\begin{equation}
	\label{eq: van der pol equations}
	\bm{\mathcal{F}}(\bm{x}) \doteq \begin{bmatrix} x_5 \\ -\omega^2 x_4 + x_5 (x_2 - x_1) - x_5 \frac{x_3 x_4^2}{1 + (x_3 / x_2)x_4^2} \end{bmatrix}, \qquad
	\bm{\mathcal{H}}(\bm{x}) \doteq x_4 + \cos(x_4), \qquad 
	\bm{x} = 
	\left[
	\begin{array}{c}
	\bm{\mu} \\ 
	\hline 
	\bm{u} 
	\end{array}
	\right]
	\doteq 
	\left[
	\begin{array}{c}
	x_1 \\
	x_2 \\
	x_3 \\
	\hline
	x_4 \\
	x_5
	\end{array}
	\right]
\end{equation}
where $\omega \doteq 240\pi$. The first three entries of $\bm{x}$ correspond to the parameters $\bm{\mu} \equiv (x_1,x_2,x_3)^T \in \R^3$, while the last two entries correspond to the state $\bm{u} \equiv (x_4,x_5)^T \in \R^2$. The sensors record the first state variable corrupted by a cosine bias term, that is $\bm{b}(\bm{x}) \doteq \cos(x_4)$. Observations are recorded once every $5$ time steps. After an initial spin-up phase, the bias model is trained on synthetic data obtained by running the transition operator on an ensemble of initial conditions. This strategy is widely adopted in the \ac{rc} community, as it allows the \ac{esn} modelling the bias to be trained so that it is largely independent of the initial conditions, thus achieving better generalisation capabilities. The training is performed using the recycle validation strategy proposed in \cite{Novoa2023,novoa2022realtime}. Following the training is the washout -- another technique frequently encountered in the context of \acp{esn} -- whose primary purpose is to correctly initialise the network once the training is complete \cite{lukosevicius2012practical}. \\

\noindent The code for this test is reported in two separate figures: in \lst{lst:van der pol} we show the main code snippet, whereas in \lst{lst:van der pol bias} we show the code for the training of the bias model. The missing code -- primarily used to set up the problem -- is reported in Appendix \ref{apdx:van der pol}. The main novelties presented in the snippets shown here with respect to those discussed in the previous section are:
\begin{itemize}
	\item Lines $31-34$ of \lst{lst:van der pol}: an additional bias function can be supplied to the \myfont{build\_observations} method (previously discussed in Subsection~\ref{subs:usage-example}) to generate biased observations.
	\item Lines $59-69$ of \lst{lst:van der pol}: a bias-aware \ac{enkf} filter with fixed multiplicative inflation $\rho \doteq 1.005$ is defined, and the \ac{da} is performed with a regularisation parameter $\gamma \doteq 10^{-3}$ (see Eq.~\eqref{eq: bias-aware enkf map cost}).
	\item Lines $18-19$ of \lst{lst:van der pol bias}: the training data used to train the \ac{esn} is constructed so that the network is taught to predict, at any given time step, the cosine bias term of the observations at the next time step. This explains why \myfont{target\_data} is defined by shifting \myfont{train\_data} by a single time step.
	\item Lines $21-33$ of \lst{lst:van der pol bias}: an \ac{esn} with $N_r \doteq 100$ hidden units is defined, with a $\tanh$ activation function and a bias term added to the input (which is normalised) and state. The keyword \myfont{connect} is related to the number of nonzero entries of the sparse reservoir matrix, while \myfont{rng} is used to seed the random number generator for reproducibility. 
	\item Lines $35-51$ of \lst{lst:van der pol bias}: the \ac{esn} is trained with the recycle validation strategy proposed in \cite{Novoa2023,novoa2022realtime}, which is a cross-validation technique that allows the best hyperparameters of the network to be selected. The training data is augmented and regularised by means of additive Gaussian noise, which was shown in \cite{novoa2022realtime} to yield more accurate forecasting by the network. Following the example of \cite{Novoa2023}, the hyperparameters consist of the spectral radius, the input scaling, and the Tikhonov regularisation parameter. The best hyperparameters are then used to train the final model.
	\item Lines $53-77$ of \lst{lst:van der pol bias}: the washout phase is run to correctly initialise the \ac{esn} before employing it in the \ac{da} window. 
\end{itemize}
The performance of the bias-aware \ac{enkf}, along with a comparison with a bias-unaware counterpart (simply given by the inflation \ac{enkf} defined at lines $61-64$ of \lst{lst:van der pol}), is shown in \fig{fig:van der pol}. The first two plots compare the predicted values of $x_1$ and $x_4$, respectively, and show how the bias-unaware option is unable to recover the true solution. This is expected, since the biased observations will never allow the filter to converge to a zero-mean Gaussian distribution for the innovation (as demonstrated by the fourth plot), which is a fundamental assumption of \acp{kf}. On the other hand, the bias-aware model recovers accurate estimates both in the parameter-state space (first two plots) and in the observation space (third plot). In particular, the predicted values of $x_4$ and the observations perfectly overlap with the true values. The fourth plot shows that the bias-aware filter accurately recovers the Gaussian assumption \ac{pdf} for the innovation.

\subsection{Discovering the Reynolds number of a turbulent Navier-Stokes flow}
\label{subs:navier stokes}

\begin{figure}[t]
  \centering
  \includegraphics{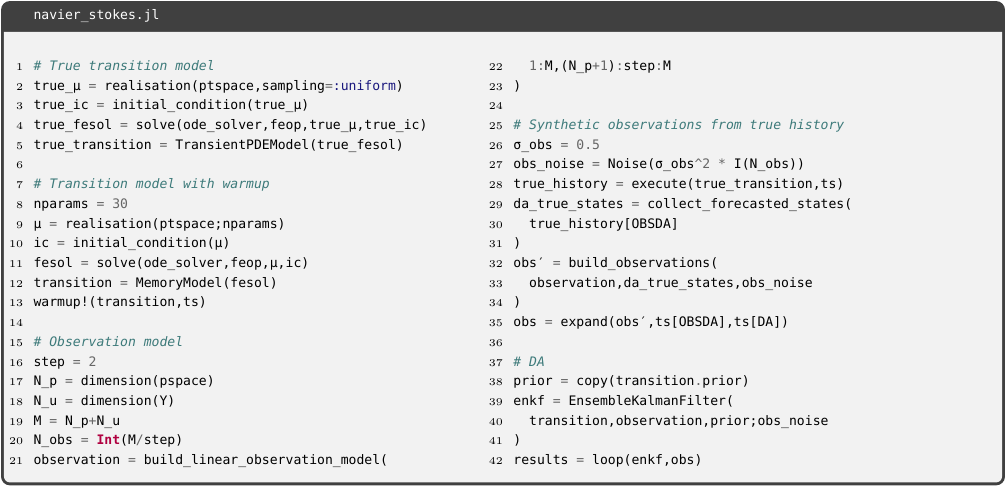}
  \caption{Code snippet used to run the \ac{da} for the Navier-Stokes example.}
  \label{lst:navier stokes}
\end{figure}

\begin{figure}[t]
	\centering
	{\legenddot{black}~True\quad\legenddot{plotblue}~\ac{enkf}}\par\smallskip
	\includegraphics[scale=0.265]{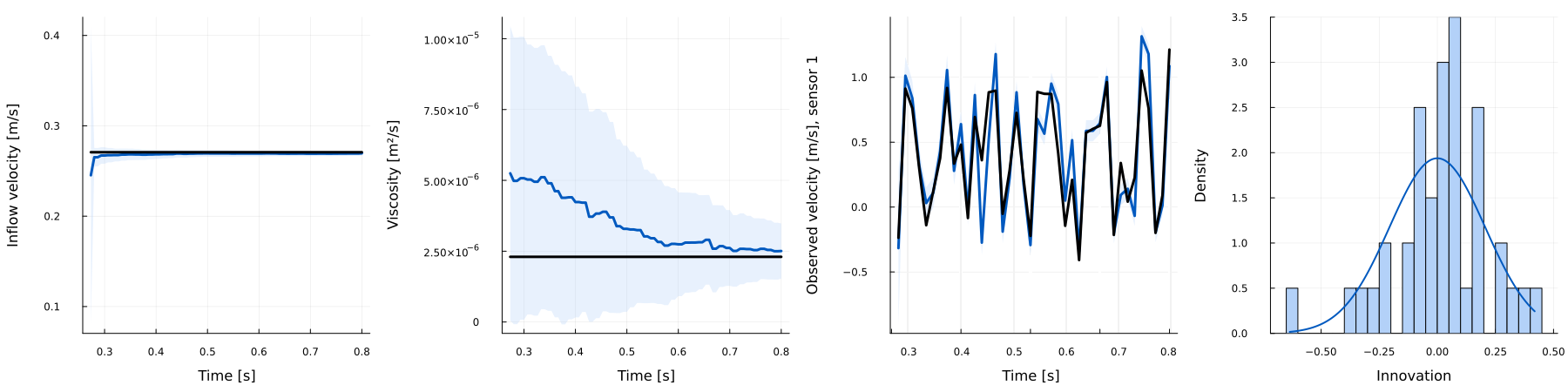}
	\caption{Navier-Stokes example. From left to right: (i) estimation of $x_1 \equiv u_{\mathrm{in}}$ in the last $0.5s$ of the assimilation window, (ii) estimation of $x_2 \equiv \nu$ in the last $0.5s$ of the assimilation window, (iii) estimation of observations in the last $0.5s$ of the assimilation window, and (iv) estimated probability density function of the innovation.}
	\label{fig: navier stokes}
\end{figure} 

\begin{figure}[t]
	\includegraphics[scale=0.4]{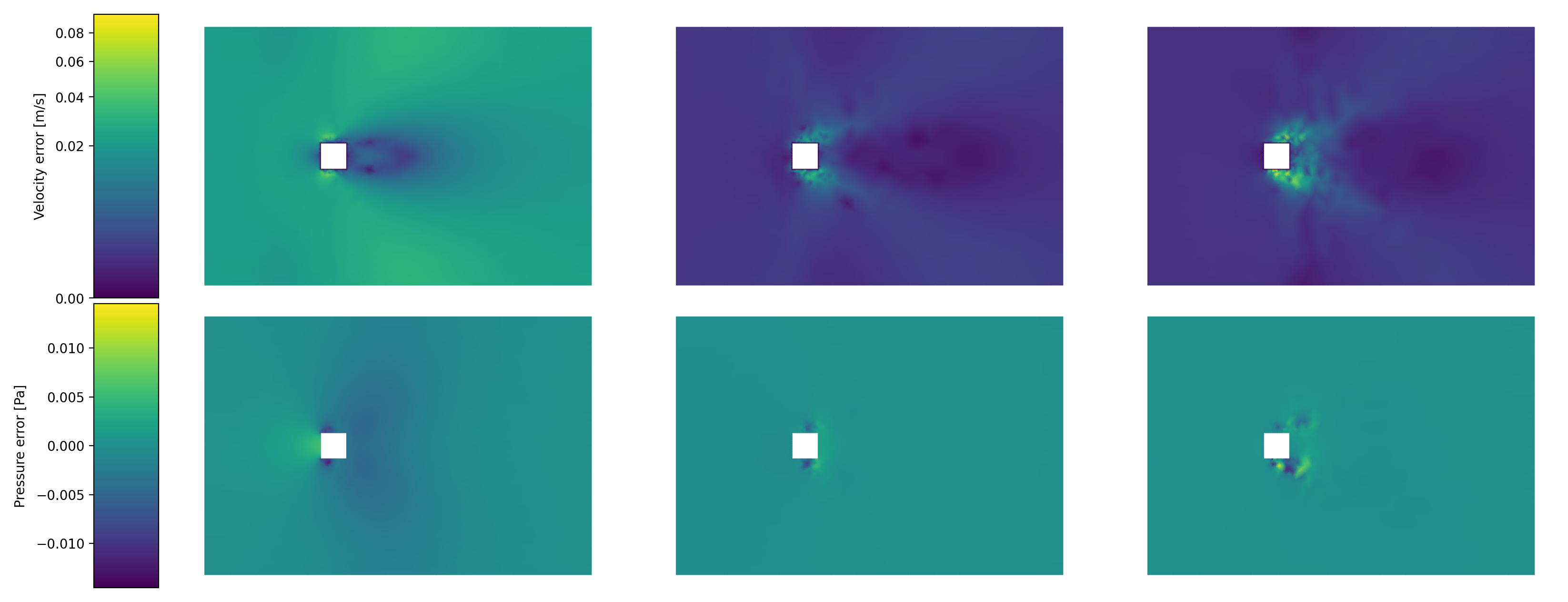}
	\caption{Velocity (top) and pressure (bottom) errors for the Navier-Stokes test at three time steps. From left to right: $t \doteq t^{20}$, $t \doteq t^{40}$, and $t \doteq t^{60}$.}
	\label{fig: navier stokes error}
\end{figure} 

In this example, we consider an \ac{enkf} procedure for the parameter-state estimation of an incompressible turbulent Navier-Stokes flow around a square cavity in a 2D rectangular domain. We consider homogeneous no-slip conditions on the top and bottom walls, a constant inflow profile on the inlet (left wall), a no-penetration condition on the outlet (right wall), and an initial condition obtained from the solution of a (steady-state) Stokes equation at $t = 0$. We add a SUPG stabilisation term to the momentum equation, whose weak formulation therefore reads as:
\begin{align}
	\label{eq: navier stokes equations}
  \int_{\Omega} \frac{d\bm{u}}{dt} \bm{v} &+ \int_{\Omega} (\bm{u} \cdot \nabla) \bm{u} \bm{v} + \int_{\Omega} \nu \nabla \bm{u} : \nabla \bm{v} - \int_{\Omega} p (\nabla \cdot \bm{v}) + \int_{\Omega} (\nabla \cdot \bm{u}) q + \int_{\Omega} \tau (\bm{u} \cdot \nabla) \bm{v} \cdot \bm{r}, \\ 
  \bm{r} &= \frac{d\bm{u}}{dt} + \nabla \cdot (\nu \nabla \bm{u}) + (\bm{u} \cdot \nabla) \bm{u} - \nabla p.
\end{align}  
The SUPG stabilisation term is given by the last integral in \eqref{eq: navier stokes equations}, where $\tau$ is a stabilisation parameter that depends on the local mesh size and the flow velocity \cite{brooks1982streamline}. The parameters of the considered problem are the viscosity and the inflow velocity in the horizontal direction, i.e. $\bm{\mu} \equiv (u_{\mathrm{in}},\nu)^T \in \R^2$. We consider a parameter space $(0.1,0.4) \times (10^{-6},10^{-5})$, and since the characteristic length of the domain is $L = 0.04$, the Reynolds number lies in the range $(400,16000)$. A P2-P1 pair of Taylor-Hood \acp{fe} is employed for the spatial discretisation, whereas a generalised $\alpha$-method is used for the time marching. At every time step, a Newton-Raphson method is used to solve the linearised system. We refer to Appendix \ref{apdx:navier stokes} for the script used to set up the experiment for additional details. This benchmark is run using a standard \ac{enkf} over $N_t \doteq 80$ \ac{da} time steps, after executing $40$ warmup iterations. A sensor records every other state variable (velocities and pressures only), and observation measurements occur every other time step, as shown in \fig{fig:stencil}. \\ 

\noindent Though this application is simpler than the previous one from a purely \ac{da} perspective, it still remains challenging due to its highly nonlinear nature. Even a slight deviation from the true parameters or states may lead the Newton solver to diverge, and thus the entire filter. For this reason, we remark that too big a departure of the initial ensemble from the true values may result in filter divergence, and thus must be chosen with care. As shown in Figs.~\ref{fig: navier stokes}-\ref{fig: navier stokes error}, the \ac{enkf} is able to overcome these challenges and accurately recover the true values of the parameters, states and observations. Notably, the fourth plot shows that the innovation is accurately modelled by a zero-mean Gaussian distribution, which confirms that the assumptions of the \ac{enkf} are satisfied. The results demonstrate the aptness of our framework in estimating the hidden Reynolds number of a turbulent flow. Though the dimension of the state ($\sim 6500$ \acp{dof}) is not intractable, the computational cost of this problem remains high, and the use of a model surrogate, though advisable, is not straightforward due to the highly nonlinear nature of the problem. 

\subsection{Data assimilation with model surrogate for the Heat equation}
\label{subs:heat equation}

\begin{figure}[t]
  \centering
  \includegraphics{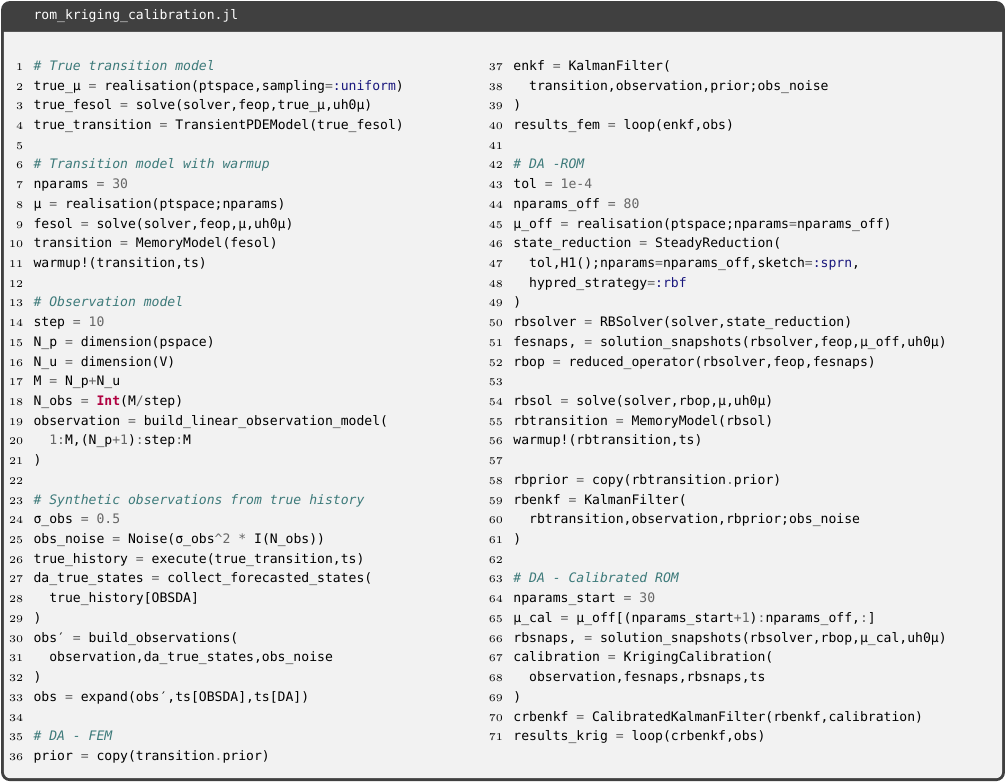}
  \caption{Code snippet used to run the \ac{da} for the Heat equation example.}
  \label{lst:heat equation}
\end{figure}

\begin{figure}[t]
  \centering
  {\legenddot{black}~True\quad\legenddot{plotred}~\ac{fe}\quad\legenddot{plotgreen}~ROM\quad\legenddot{plotblue}~Calibrated ROM}\par\smallskip
  \includegraphics[scale=0.27]{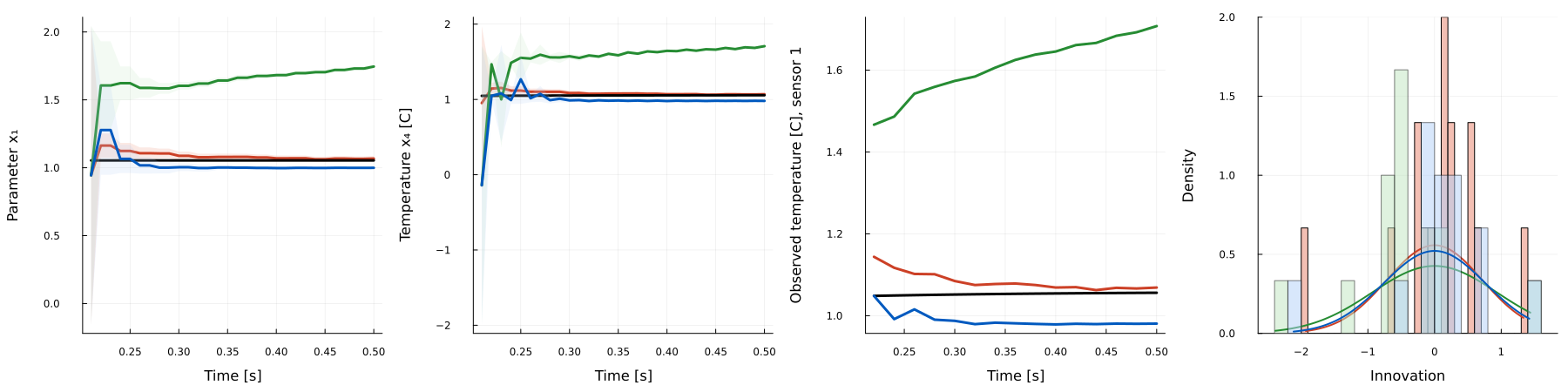}
  \caption{Heat equation example. From left to right: (i) estimation of $x_1 \equiv \mu_1$ in the last $0.5s$ of the assimilation window, (ii) estimation of $x_4 \equiv u_1$ in the last $0.5s$ of the assimilation window, (iii) estimation of observations in the last $0.5s$ of the assimilation window, and (iv) estimated probability density function of the innovation.}
  \label{fig: heat equation}
\end{figure}

\begin{figure}[t]
	\includegraphics[scale=0.4]{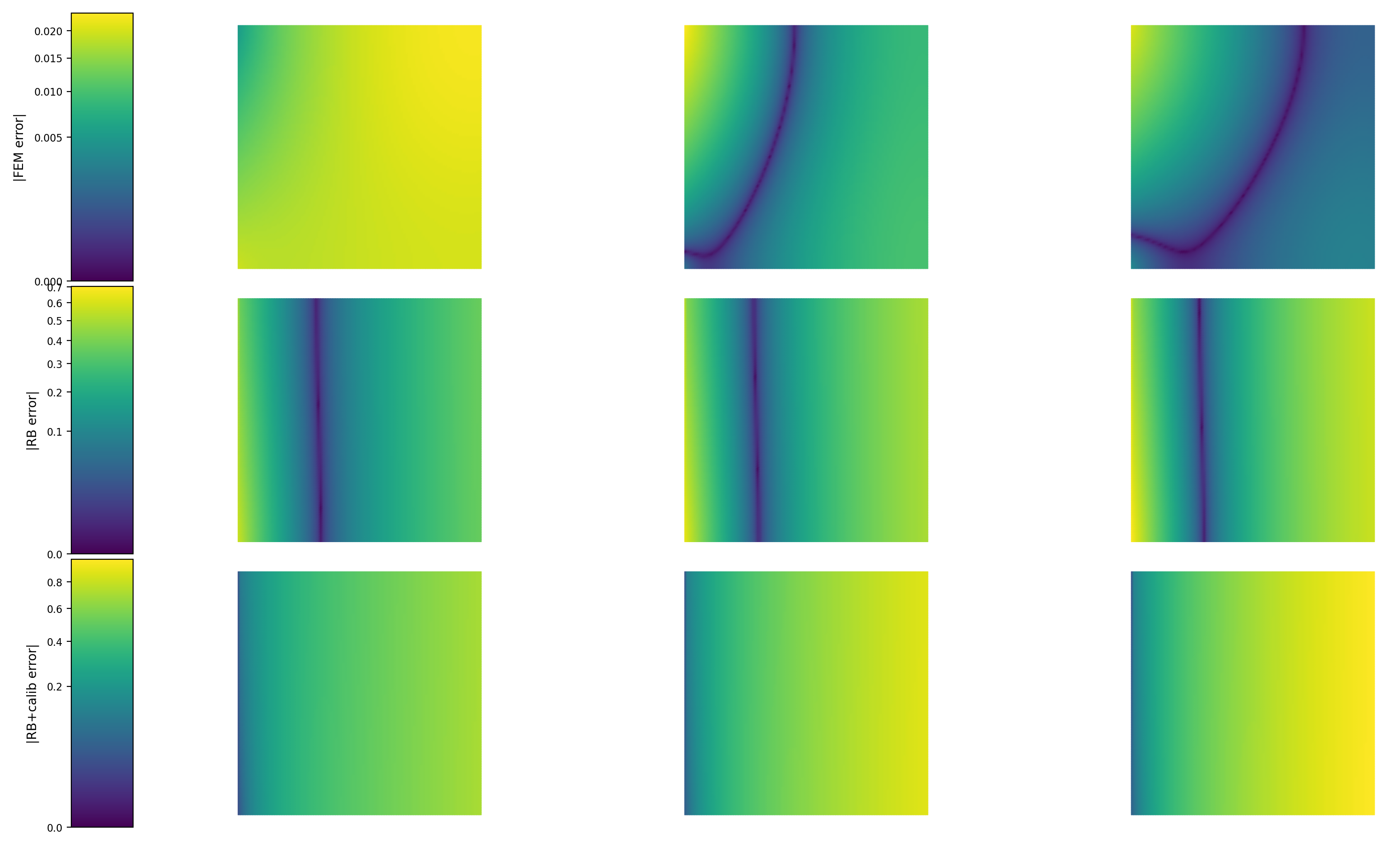}
	\caption{Heat equation errors at three time steps. From left to right: $t \doteq t^{10}$, $t \doteq t^{20}$, and $t \doteq t^{30}$.}
	\label{fig: heat equation error}
\end{figure} 

In this example, we consider an \ac{enkf} procedure for the parameter-state estimation of a Heat equation on a unit square. A comparison is made between \ac{da} loops obtained with transition operators given by (i) the \ac{fe} method, (ii) a \ac{rb} surrogate, and (iii) a \ac{rb} surrogate with the kriging calibration described in Subsection~\ref{subs:rem}. We consider non-homogeneous Dirichlet boundary conditions on the left wall, non-homogeneous Neumann boundary conditions on the right wall, and homogeneous Neumann boundary conditions on top and bottom walls. The problem's weak formulation reads as:
\begin{equation}
	\label{eq: heat equation}
	\int_{\Omega} \frac{du}{dt} v + \int_{\Omega} \alpha \nabla u \cdot \nabla v = \int_{\Omega} v, \qquad 
	u = u_{\textsc{D}} \quad \text{on} \quad \{0\} \times (0,1), \qquad 
	-\alpha\nabla u \cdot \bm{n} = u_{\textsc{N}} \quad \text{on} \quad \{1\} \times (0,1),
\end{equation}
where 
\begin{equation*}
	\alpha(\bm{x};t,\bm{\mu}) = 1 + \exp \left(-\sin \left(t\right)^2 x_1 / \sum_i \mu_i\right),
	\qquad
	u_{\textsc{D}}(\bm{x};t,\bm{\mu}) = \mu_1 \exp \left(-x_2 / \mu_3\right),
	\qquad
	u_{\textsc{N}}(\bm{x};t,\bm{\mu}) = \left|\cos \left(t / \mu_2\right)\right|.
\end{equation*}
The parameters of the considered problem are $\bm{\mu} \equiv (\mu_1,\mu_2,\mu_3)^T \in \R^3$, which are constrained to lie in the parametric domain $(1,10)^3$. A P1 \ac{fe} is employed for the spatial discretisation, whereas a Backward Euler integrator is used for time integration. The problem is initialised from a homogeneous initial condition at $t = 0$. Observations are recorded at every other time step, and one sensor is placed at every $10$ spatial \acp{dof}. The core of the parameter-estimation procedure is illustrated in \lst{lst:heat equation}, whereas the missing lines of code are reported in the Appendix \ref{apdx:heat equation}. This experiment is run using a standard \ac{enkf} over $N_t \doteq 30$ \ac{da} time steps, after executing $20$ warmup iterations. \\ 

\noindent The results of the \ac{da} procedure are shown in Figs.~\ref{fig: heat equation}-\ref{fig: heat equation error}. From Fig.~\ref{fig: heat equation}, it appears that the calibrated \ac{rom} greatly outperforms the standard \ac{rom} in terms of accuracy. Indeed, the latter seems to not converge to the right values, due to the relatively high \ac{rom} errors with respect to the observation errors, which cause the analysis step to correct to a wrong value. However, Fig.~\ref{fig: heat equation error} depicts a more unclear picture, as it shows the calibrated filter does not uniformly outperform the standard \ac{rom} in terms of accuracy in all state variables. From this, it appears that the second plot of Fig.~\ref{fig: heat equation} demonstrating how well the calibrated \ac{rom} estimates $x_4$ compared to the plain \ac{rom} is possibly a ``lucky'' shot. Indeed, $x_4$ is the temperature value recorded at the point $(0.01,0)$, which appears to be in a region of the domain where the calibrated \ac{rom} error is relatively low. Interestingly, the same is true for the parameters: though we empirically verified that parameters are generally better captured by the calibrated \ac{rom}, we have also simulated scenarios where some parameters are better estimated by the standard \ac{rom}. This is not incompatible with the findings of \cite{pagani2017efficient}, which show that the kriging calibration is able to reduce the \ac{rom} error in the observation space, but not necessarily in the parameter-state space. A better option to improve the accuracy of the \ac{rom} surrogate would have been, in this instance, to increase the accuracy of the \ac{rom} itself by increasing the number of \ac{rb} basis functions, rather than relying on the kriging calibration. These findings are a motivation to explore alternative \ac{rom} error-aware methods in future work. In any case, \lst{lst:heat equation} demonstrates how easy it is to implement \ac{da} loop with surrogate models, both calibrated and standard. Lastly, we remark that the \ac{rom} surrogate is about $15$ times cheaper than the \ac{fe} model in memory ($13.56$ Gb for the \ac{fe}, $0.91$ Gb for the \ac{rom}), whereas the calibrated \ac{rom} is $\approx 2.7$ times cheaper; these significant speed-up measures demonstrate the usefulness of \ac{rom} surrogates in \ac{da} frameworks.
\section{Conclusions}
\label{sec:conclusions}
In this work, we introduce Opal, a Julia package that provides a unified framework for sequential and variational inference for a variety of dynamical systems. The package is designed to be composable, allowing users to build complex inference tools by combining basic methods with advanced capabilities. Opal integrates natively with the SciML ecosystem for \ac{ode}-governed systems and with Gridap/GridapROMs for both full-order and reduced-order discretisations of PDEs. We demonstrate the library on several examples, including a Lorenz-63 benchmark, a Van der Pol oscillator problem with biased observations, a turbulent Navier-Stokes flow past a square cavity, and a heat equation using a model surrogate. The results show that Opal can effectively estimate hidden states and uncertain parameters from partial, noisy, and potentially biased observational data. These numerical experiments are challenging, as they combine a wide range of concepts and implementation building blocks, such as \acp{rom}, strongly nonlinear saddle-point dynamics, and \ac{rc}-based correction of structural model errors. \\

\noindent We envision three main directions for future work. First, we plan to include multi-fidelity inference methods, which can leverage the availability of multiple models of varying fidelity to improve accuracy for a given computational budget. A first step in this direction could be to insert our existing \ac{rom} frameworks within the multi-fidelity pipelines, although we may eventually develop more advanced strategies, such as recently emerging deep learning approaches \cite{ZACCHEI2026118916}. Second, we plan to develop a more comprehensive suite of \acp{rom} for \ac{pde}-governed systems, and to further strengthen the \ac{rom} error-aware strategies available in the library. Finally, we plan to extend the library to handle more complex applications, mainly in the context of hydrodynamics and fluid-structure interaction problems, which are of particular interest to us within the scope of offshore engineering.

\section*{Acknowledgments}
This publication is part of the DigiOcean4Solar project, with file number 21225, of the NWO Talent Programme Vidi AES 2023, which is financed by the Dutch Research Council (NWO) under grant ID \href{https://doi.org/10.61686/OPCTU16570}{10.61686/OPCTU16570}.

\bibliographystyle{elsarticle-num}
\bibliography{references}

\appendix

\newpage 
\section[Appendix A]{Appendix A: missing code snippet from the Van der Pol example}
\label{apdx:van der pol}

In \lst{lst:missing van der pol} we collect the code snippet omitted from the Van der Pol example for brevity.
\begin{figure}[H]
  \centering
  \includegraphics{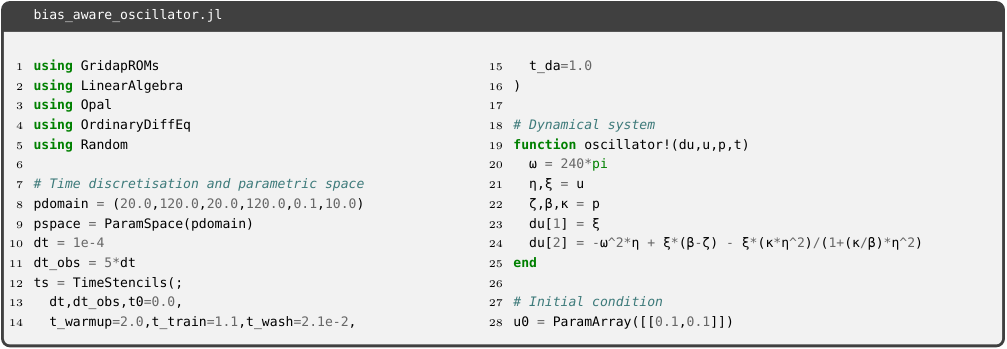}
  \caption{Missing code snippet from the Van der Pol example.}
  \label{lst:missing van der pol}
\end{figure}

\newpage 
\section[Appendix B]{Appendix B: missing code snippet from the Navier-Stokes example}
\label{apdx:navier stokes}
Although \lst{lst:missing navier stokes} appears to be quite convoluted, it follows a similar structure to \lst{lst:missing van der pol}, with the only main difference being that an initial condition is manually set as the solution of a steady-state Stokes problem at $t = 0$, and several \ac{fe}-specific quantities are introduced. This requires several additional lines of code to set up the workflow of this numerical experiment, but the underlying logic remains the same as in the previous examples.
\begin{figure}[H]
  \centering
  \includegraphics[scale=0.95]{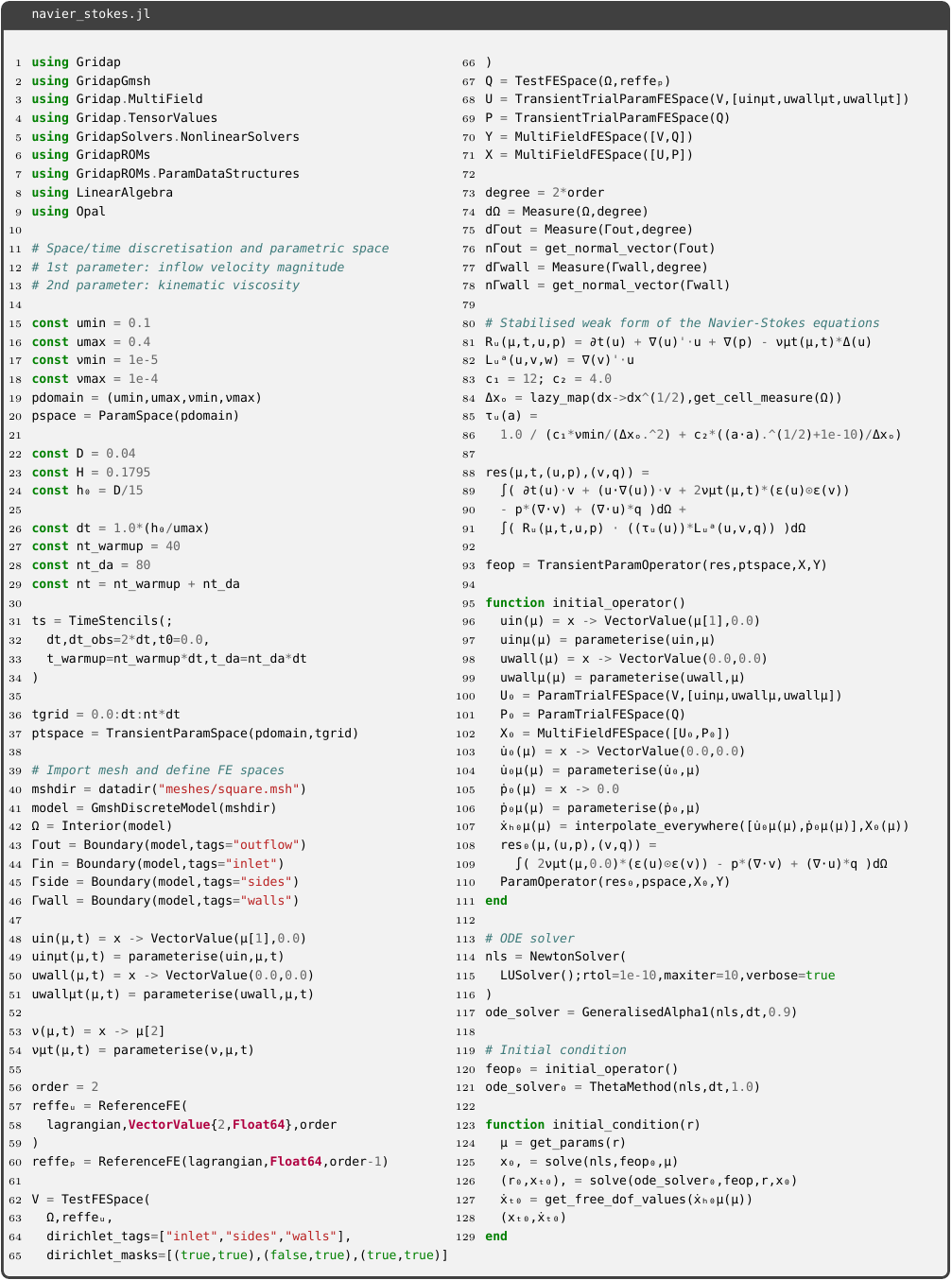}
  \caption{Missing code snippet from the Navier Stokes example.}
  \label{lst:missing navier stokes}
\end{figure}

\newpage 
\section[Appendix C]{Appendix C: missing code snippet from the Heat equation example}
\label{apdx:heat equation}
In \lst{lst:missing heat equation}, we only underline that the unknown parameters are sought in the parametric domain $(1,10)^3$, and that all transition operators are initialised from a homogeneous initial condition $u^0 = 0$ at $t = 0$.

\begin{figure}[H]
  \centering
  \includegraphics{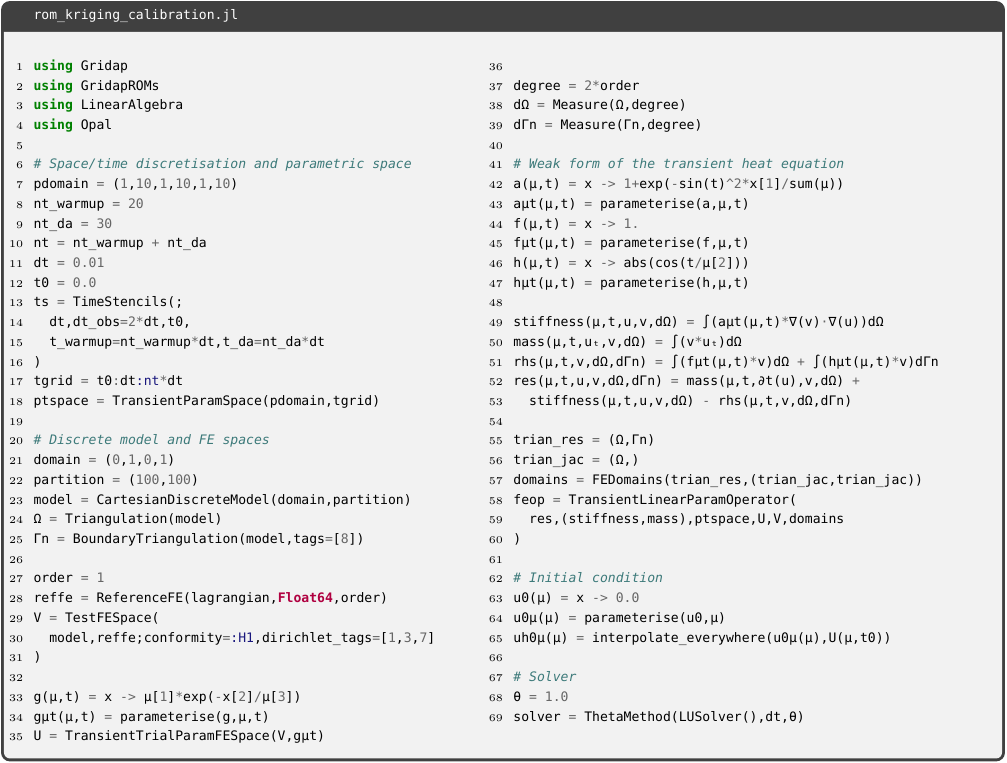}
  \caption{Missing code snippet from the Heat equation example.}
  \label{lst:missing heat equation}
\end{figure}

\end{document}